%% file: ms.tex
\documentclass[sigconf]{acmart}


\copyrightyear{2018}
\acmYear{2018}
\setcopyright{rightsretained}
\acmConference[EuroSys '18]{Thirteenth EuroSys Conference 2018}{April
23--26, 2018}{Porto, Portugal}
\acmBooktitle{EuroSys '18: Thirteenth EuroSys Conference 2018, April
23--26, 2018, Porto, Portugal}
\acmDOI{10.1145/3190508.3190551}
\acmISBN{978-1-4503-5584-1/18/04}

\usepackage{microtype}

\newcommand{\tf}{TensorFlow}  
\date{}

\begin{document}
\title{Dynamic Control Flow in Large-Scale Machine Learning}

\author{Yuan Yu}
\authornote{Work done primarily at Google Brain.}
\affiliation{Microsoft}
\email{yuanbyu@microsoft.com}

\author{Mart\'{\i}n Abadi}
\affiliation{Google Brain}
\email{abadi@google.com}

\author{Paul Barham}
\affiliation{Google Brain}
\email{pbar@google.com}

\author{Eugene Brevdo}
\affiliation{Google Brain}
\email{ebrevdo@google.com}

\author{Mike Burrows}
\affiliation{Google Brain}
\email{m3b@google.com}

\author{Andy Davis}
\affiliation{Google Brain}
\email{andydavis@google.com}

\author{Jeff Dean}
\affiliation{Google Brain}
\email{jeff@google.com}

\author{Sanjay Ghemawat}
\affiliation{Google}
\email{sanjay@google.com}

\author{Tim Harley}
\affiliation{DeepMind}
\email{tharley@google.com}

\author{Peter Hawkins}
\affiliation{Google Brain}
\email{phawkins@google.com}

\author{Michael~Isard}
\affiliation{Google Brain}
\email{misard@google.com}

\author{Manjunath Kudlur}
\authornotemark[1]
\affiliation{Cerebras Systems}
\email{manjunath@cerebras.net}

\author{Rajat Monga}
\affiliation{Google Brain}
\email{rajatmonga@google.com}

\author{Derek Murray}
\affiliation{Google Brain}
\email{mrry@google.com}

\author{Xiaoqiang Zheng}
\affiliation{Google Brain}
\email{zhengxq@google.com}

\renewcommand{\shortauthors}{Y. Yu et al.}

\begin{abstract}
Many recent machine learning models rely on fine-grained dynamic control flow for training and inference.
In particular, models based on recurrent neural networks and on reinforcement learning depend on
recurrence relations, data-dependent conditional execution, and other features that call for dynamic control flow.
These applications benefit from the ability to make rapid control-flow decisions across a
set of computing devices in a distributed system.
For performance,
scalability, and expressiveness, a machine learning system must
support dynamic control flow in distributed and heterogeneous environments.

This paper presents a programming model for distributed machine learning
that supports dynamic control flow.
We describe the design of the programming model, and its
implementation in {\tf}, a distributed
machine learning system.
Our approach extends the use of dataflow graphs
to represent machine learning models, offering
several distinctive features. First, the branches of conditionals and
bodies of loops can be partitioned across many machines to run on a
set of heterogeneous devices, including CPUs, GPUs, and custom ASICs. Second,
programs written in our model support automatic differentiation and distributed
gradient computations, which are necessary for training machine learning models that
use control flow. Third, our choice of non-strict semantics enables multiple loop
iterations to execute in parallel across machines, and to
overlap compute and I/O operations.

We have done our work in the context of {\tf}, and it has been used extensively in
research and production. We evaluate it using several real-world applications,
and demonstrate its performance and scalability.

\end{abstract}

\maketitle

\input{intro}
\input{prog}
\input{arch}
\input{impl}
\input{autodiff}

\input{eval}
\input{rw}

\input{conc}

\begin{acks}
We gratefully acknowledge contributions from our colleagues at Google and from
members of the wider machine learning community, and the feedback that we have
received from them and from the many users of {\tf}. In particular, Skye
Wanderman-Milne provided helpful comments on a draft of this paper. We also
thank our shepherd, Peter Pietzuch, for his guidance in improving the paper.
\end{acks}

\bibliography{controlflow}
\bibliographystyle{acm}

\end{document}

%% file: intro.tex
\section{Introduction}\label{sec:intro}

Advances in machine learning and their many
applications have brought into focus conflicting design objectives for
the underlying systems. These systems should be scalable:
they should use hardware resources efficiently, on platforms from
individual phones to powerful datacenters that comprise
CPUs, GPUs, and custom ASICs such as TPUs~\cite{tpu-blogpost,tpu-isca}.
At the same time,
systems should be expressive and flexible to support
both production and research.

\begin{figure}
  \begin{center}
  \includegraphics[width=0.9\linewidth]{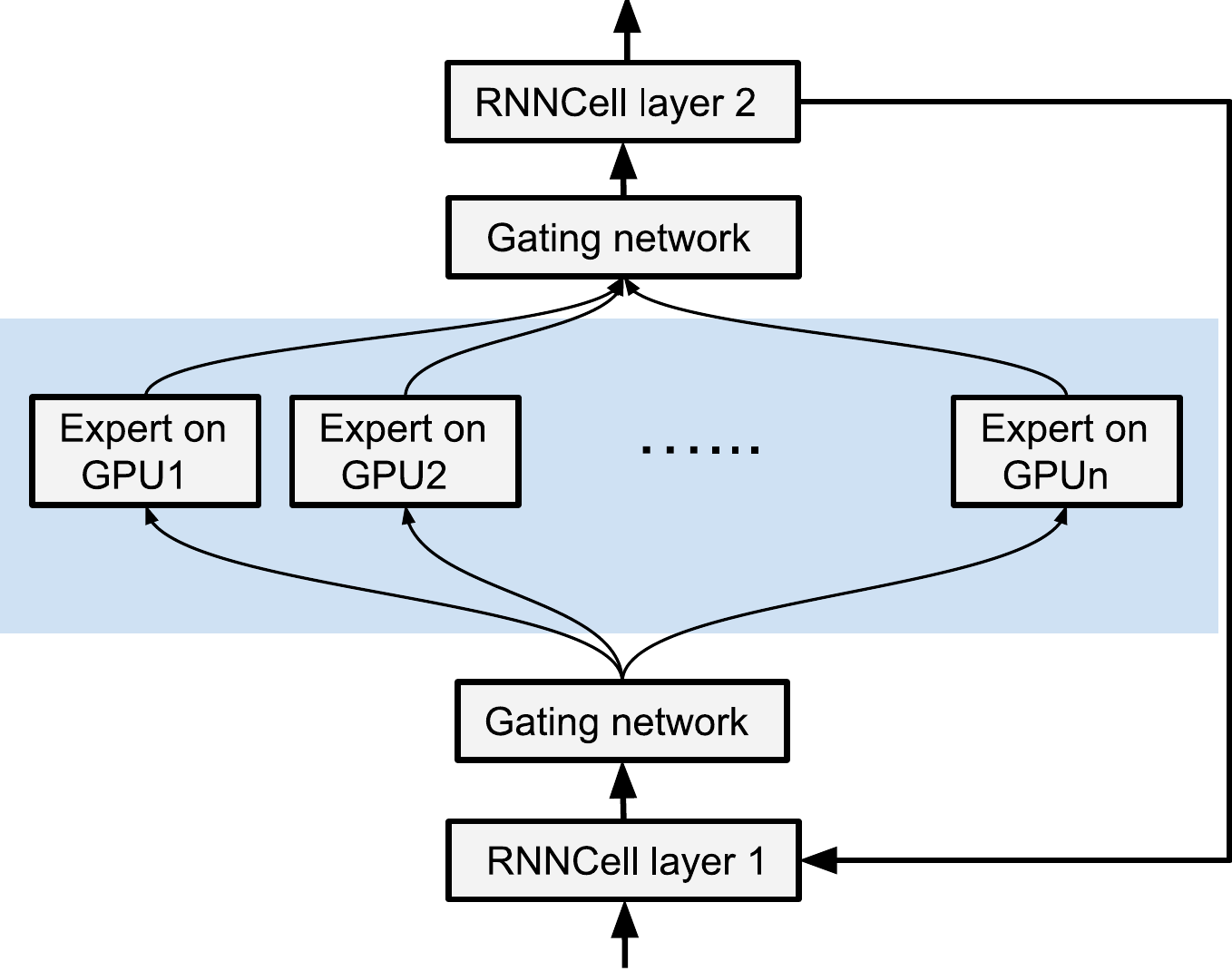}
\end{center}
\caption{An example model architecture.}\label{fig:intro:moe}
\end{figure}

As an example, consider the model architecture depicted in
Figure~\ref{fig:intro:moe}. It is a simplified version of one
employed at Google for language processing. This architecture
includes two RNN (Recurrent Neural Network) layers, and a ``mixture of
experts'' (MoE) that provides dynamic, learned connections between
the RNN layers.  While simple feedforward neural networks are akin to
straight-line code, this architecture makes essential use of dynamic
control flow, with all the difficulties that it entails.  An
embodiment of the resulting system may run over a distributed
infrastructure, with many components on different devices.  Implementation
choices can have a large impact on the performance of such a system
and its memory requirements.  For instance, RNNs give rise to
challenging trade-offs between memory footprint and computation time
(e.g.,~\cite{DBLP:journals/corr/GruslysMDLG16}).
Distributing computations over multiple GPUs and swapping tensors
between GPU and host memory can alleviate these memory constraints, and adds
another dimension to those trade-offs.

Both the building blocks of machine learning (e.g., individual cells in
Figure~\ref{fig:intro:moe}) and the architectures built up using these blocks
have been changing rapidly. This pace appears likely to continue. Therefore,
rather than defining RNNs, MoEs, and other features as primitives of a
programming model, it is attractive to be able to implement them in terms of
general control-flow constructs such as conditionals and loops. Thus, we
advocate that machine learning systems should provide general facilities for
dynamic control flow, and we address the challenge of making them work
efficiently in heterogenous distributed systems consisting of CPUs, GPUs, and
TPUs.




Modern machine learning frameworks typically use data\-flow graphs to
represent computations, and a client process to drive those computations on a
collection of accelerators.
The frameworks offer two main approaches for control flow:
\begin{itemize}
\item The \emph{in-graph}
  approach~\cite{alrfou2016theano,tensorflow2016} encodes control-flow decisions
  as operations in the dataflow graph.
\item The \emph{out-of-graph}
  approach~\cite{jia2014caffe,collobert2002torch,chen2015mxnet}
  implements control-flow decisions in the client process, using
  control-flow primitives of a host language (e.g., Python).
\end{itemize}
When designing 
{\tf}~\cite{tensorflow2016} we favored the in-graph approach, because
it has advantages at both compile and run time.  With this approach,
the machine learning system processes a single unified dataflow graph, and
can perform whole-program optimization. In particular, it
can determine exactly which intermediate values from the execution of a machine
learning model are necessary to compute the gradients that
are pervasive in machine learning algorithms, and it can decide to cache those
values.
In addition, this approach allows the entire computation to stay
inside the system runtime during execution. This feature is
important in heterogeneous environments where
communication and synchronization with the client process can be
costly.

If the in-graph dynamic control-flow features of TensorFlow are not
used, a program must rely on the control-flow
features of the host language, or resort to static unrolling.
In-graph loops allow us to obtain more parallelism than out-of-graph loops.
There can be a factor of 5
between the number of iterations per second in a microbenchmark that compares
in-graph and out-of-graph loops distributed across 8 GPUs on one machine
(Section~\ref{ss:eval:micro}).
We explore and exploit these advantages in our work.


In our work, high-level control-flow constructs are compiled to dataflow graphs
that include a few simple, powerful primitives similar to those of dynamic
dataflow machines~\cite{Arvind:1986, Arvind:1990}.  These graphs can then be arbitrarily
partitioned and executed on a set of heterogeneous
devices, using a lightweight coordination mechanism. 

Since automatic differentiation is an important technique for training machine
learning models, we also show how to support it efficiently in the presence
of dynamic control flow. Given a graph, we demonstrate how to add a
subgraph that computes gradients, and develop optimizations and memory-management
techniques for this computation. If the original graph includes control-flow
constructs, the new subgraph will as well.
These subgraphs can also be partitioned and
executed on a set of heterogeneous devices.  

Our implementation allows parallelism and asynchrony, providing a
non-strict semantics: operations in a conditional branch or a loop iteration can
run as soon as their inputs are available. This
property enables us to overlap the execution of
control-flow logic on CPU, compute kernels on GPU, and memory-copy
operations between CPU and GPU. 

While it is difficult to measure ease-of-use or flexibility, we
have evidence that our design and implementation are suitable for a
variety of real-world needs.  Our code is part of the core
{\tf} distribution, and is used widely in production and research.
We analyzed more than 11.7
million unique graphs for machine learning jobs at Google
over the past year, and found that approximately 65\%
contain some kind of conditional computation, and
approximately 5\% contain one or more loops. Teams that conduct
research on model
structures for machine learning, and several open-source projects
(e.g., SyntaxNet~\cite{syntaxnet},
TensorFlow Fold~\cite{tensorfold}, and Sonnet~\cite{sonnet}) rely
on our work.
Accordingly, our evaluation of performance and scalability
reflects results for real applications in the context of a real system.

The original
TensorFlow paper~\cite{tensorflow2016} briefly sketched our approach to dynamic
control flow, but did not provide a detailed design or evaluation. The present
work goes much further in both respects. To the best of our knowledge, there is no previous dataflow system that supports both distributed control flow and automatic differentiation.

In sum, the main contributions of our work are:
\begin{itemize}
\item A design for how to incorporate in-graph dynamic control flow in machine learning,
  including,
  in particular, automatic differentiation.

\item A corresponding implementation that allows parallel and distributed
  execution across CPUs, GPUs, and custom machine learning accelerators.

\item An evaluation that characterizes the performance and scalability of our techniques, and analyzes the impact of various choices.

\item Extensive experience with users, which gives evidence of the expressiveness and flexibility of the system.
\end{itemize}

Section~\ref{sec:prog} discusses programming constructs for
control flow and their applications in machine learning. Section~\ref{sec:arch}
provides an architectural overview, highlighting challenges related to dynamic
control flow. Section~\ref{sec:impl} describes our design and implementation.
Section~\ref{sec:diff} considers two critical aspects of the system:
automatic differentiation and memory management. Section~\ref{sec:eval}
evaluates the performance and scalability of the system. Section~\ref{sec:rw}
discusses related work.


%% file: prog.tex
\section{Programming with Control Flow}\label{sec:prog}

In this section, we briefly review {\tf}'s programming model
and describe its support for dynamic control flow.  We also discuss how dynamic
control flow is being used in machine learning.


\subsection{Programming Interface}\label{ss:prog:interface}

{\tf} employs a two-level programming model: programmers construct a dataflow graph using a
high-level programming language; and the {\tf} runtime takes a complete dataflow graph,
optimizes it, and executes it on a set of devices. {\tf}
provides language bindings for many programming
languages including Python, Java, and Go. In this paper, we use Python
for illustration.

To simplify the task of constructing machine learning models, {\tf}'s API
exposes a large collection of pre-defined operators.
The operators are embedded as functions in the host programming
languages. They support a wide range of mathematical operations on
\emph{tensors}, which are dense multi-dimensional arrays of
basic data types such as float, integer, and string.
The operators also include control-flow constructs, based on the work described
in this paper. The most basic ones are
\texttt{cond} and \texttt{while\_loop}, for expressing
conditional and iterative computation, respectively:
\begin{itemize}
  \item
\texttt{cond(pred, true\_fn, false\_fn)} represents a conditional
computation, where \texttt{pred} is a boolean tensor, and \texttt{true\_fn} and
\texttt{false\_fn} are functions that construct the
subgraphs for the respective branches. Both \texttt{true\_fn} and
\texttt{false\_fn} return a tuple of tensors, with
matching data types for each component; the result of \texttt{cond} is a tuple of
tensors, representing the result of the branch that executes.

\item
\texttt{{\nolinebreak[4]while\_loop}(pred, body, inits)} represents
an iterative computation,
where \texttt{pred} and \texttt{body} are functions that
construct the subgraphs for the loop
termination condition and the loop body; \texttt{inits} is a tuple
of tensors that specifies the initial values of the loop variables. Both \texttt{pred}
and \texttt{body} take a tuple of loop variables as arguments; \texttt{pred} returns a boolean
tensor, and \texttt{body} returns a tuple of updated loop variables.
\end{itemize}
Other control-flow constructs include higher-order functions such as
\texttt{map\_fn},
\texttt{foldl}, \texttt{foldr}, and \texttt{scan}. However, the number of
primitives remains small: the higher-order functions are actually defined
in terms of \texttt{\nolinebreak[4]while\_loop}. We support automatic differentiation
for all these operators.

{\tf} provides a few common data structures,
including mutable variables and queues, and---as in conventional programs---data
structures are an important tool for organizing a dynamic computation.
%
%
%
%
%
%
{\tf} initially lacked support for arrays of tensors
with random read/write accesses, and we therefore augmented it with a new type
of TensorArray objects.
Unlike other data structures, TensorArrays can be used to store values
consumed and produced by loops in a differentiable way, and they are a
useful complement to loops.
The basic operations on a TensorArray are
\texttt{ta.read(ix)}, which reads the array element at index \texttt{ix}, and
\texttt{ta.write(ix, t)}, which writes the value \texttt{t} at index
\texttt{ix}. A tensor may be converted to or from a TensorArray using the
\texttt{ta.unstack()} and \texttt{ta.stack()} methods, respectively.
We also
support automatic differentiation for these operations
(Section~\ref{sec:diff:tensorarray}). TensorArrays are now used
extensively in advanced machine learning models.

As an illustration of the expressiveness of our primitives,
Figure~\ref{fig:prog:scan} shows how we can define a higher-order \texttt{scan},
or generalized prefix-sum function, in terms of TensorArray objects and a
\texttt{while\_loop()}.
Given a tensor \texttt{elems} with dimensions \texttt{[n, \ldots]}, and an
initial value \texttt{init}, \texttt{scan} computes a new tensor that contains
the values of the expressions:
\texttt{fn(init, elems[0,\ldots])},
\texttt{fn(fn(init, elems[0,\ldots]), elems[1,\ldots])},
\textellipsis.
It exemplifies
a common computation pattern in machine learning:
a tensor \texttt{elems} is first unstacked into a TensorArray of subtensors,
\texttt{elem\_ta}; a function \texttt{fn} is then
repeatedly applied to those subtensors; and finally the results are
packed back into a single tensor.

\begin{figure}
\begin{verbatim}
def scan(fn, elems, init):
  elem_ta = 
    TensorArray(dtype=elems.dtype).unstack(elems)
  result_ta = TensorArray(dtype=init.dtype)
  n = elem_ta.size()
  def pred(i, a, ta):
    return i < n
  def body(i, a, ta):
    a_out = fn(a, elem_ta.read(i))
    ta = ta.write(i, a_out)
    return (i + 1, a_out, ta)
  _, _, r = 
    while_loop(pred, body, (0, init, result_ta))
  return r.stack()
\end{verbatim}
\caption{The \texttt{scan} operator can be defined using \texttt{while\_loop} and TensorArrays.}\label{fig:prog:scan}
\end{figure}

\subsection{Control Flow in Machine Learning}\label{ss:prog:apps}

Most traditional neural networks (including multi-layer perceptrons
and convolutional neural networks) were static, in the sense that they
were composed of a fixed number of layers, each with fixed operators.
For learning over sequences (in particular, with RNNs),
fixed-length iterations would typically be unrolled statically.
In cases where some dynamic behavior was desired, the typical solution
was to use client-side control-flow decisions.

However, for the past few years, we have seen growing demand for
dynamic control flow, especially in applications of
recurrent and tree neural networks~\cite{DBLP:journals/corr/LooksHHN17, DBLP:journals/corr/AndorAWSPGPC16, DBLP:journals/corr/Graves16} and of reinforcement
learning~\cite{mnih2015human}. This section gives a few examples.

Beyond its programming-model implications, this trend raises the bar for
implementations. For example, dynamic RNN models may operate over
sequences of thousands of inputs. Since memory usage often grows linearly with
sequence length, the amount of memory available on an accelerator is often the
key factor that limits sequence length. Thus, this trend implies that
memory optimizations are increasingly important.
We address these challenges in later sections.

\paragraph{RNNs}
The basic computation pattern of an RNN is to apply a \textit{cell} function
to every element of a sequence. A cell function takes as input a sequence
element and the current ``state'' value; it returns an output sequence
element and the updated state value, which is used when processing the next
element.
RNNs are thus well-suited for
processing sequential data; some of their variants can also be applied to trees and other unbounded data structures.

RNNs are widely used in machine learning.
One important application of RNNs is
Neural Machine Translation (NMT)~\cite{Sutskever-et-al-NIPS2014, wu2016gnmt}, for which a model is composed of an
encoder and a decoder. Both the encoder and decoder can conveniently be expressed as while-loops
on variable-length sentences. We implemented the \texttt{dynamic\_rnn} operator in {\tf}
using while-loops and TensorArray objects.

An RNN cell can be very compute-intensive, so we often need to
distribute its execution across multiple devices.
Some recent
advances in RNN models have made the cell function itself perform a dynamic computation. For
example, Adaptive Computation Time~\cite{DBLP:journals/corr/Graves16} uses a nested while-loop to
learn how many computational steps to take at each time step of the
outer loop of the RNNs. This application exercises our support for distributed
execution of nested while-loops, and for their automatic differentiation.

Another example is the inclusion of an MoE layer
inside the RNN cell~\cite{DBLP:journals/corr/ShazeerMMDLHD17}. Such a layer
typically comprises a set of \textit{experts}---simple neural networks that are
specialized for a particular prediction task---and a gating function that
chooses which experts to consult for a particular example. The experts and the
gating function are subnetworks with trainable parameters. Since there can be a
large number of experts (e.g., a few thousand in some NMT models) each typically with
1 million parameters, the experts are typically distributed across many
machines.


\paragraph{Reinforcement learning}  Reinforcement learning is a form of machine
learning where an agent interacts with its environment by performing a sequence
of actions according to some learned policy and receives rewards either at each
interaction or after a sequence of interactions.  The agent's goal
is to choose actions that maximize the sum of these rewards.

Unlike in standard supervised learning, the agent's actions and the resulting
rewards need not be deterministic or differentiable, so traditional
backpropagation is insufficient and additional training techniques are
required.  These techniques generally benefit from dynamic
control flow.

For example, some of our users write in-graph while-loops in which the agent
interacts with its environment, repeatedly, and at each step the agent's
actions are chosen by sampling according to learned probabilities.  The
expectation on the total rewards is a differentiable function of those
probabilities, so can be optimized by gradient descent. Other users employ
in-graph conditionals to dynamically read or write agent experiences to an
in-graph database in parallel to the generation of the agent's actions. Some
users also employ in-graph conditionals to create agents that choose whether to
act randomly (explore) or act according to a learned policy (exploit).

\paragraph{Other usage}

In addition to these examples of applications that emphasize machine learning
architectures, we have seen several more mundane but useful applications of
dynamic control flow.
For instance, some of our users have relied on in-graph while-loops
for programming training loops. (Generally, training loops are defined
out-of-graph, in the host high-level programming language.)
In this use case, a single coordinator process controls
many workers that may not even be located in the same datacenter; in-graph
control flow allows workers to make progress on training
independently, without synchronizing with the coordinator between steps.
Others
have relied on in-graph conditionals for doing updates selectively,
for example updating model parameters only when updates are
sufficiently large, or updating only some model parameters in certain
training steps.

%% file: arch.tex
\section{System Architecture}\label{sec:arch}

In this section we review {\tf} as a representative system architecture for a
modern machine learning system~\cite{tensorflow2016}, and discuss how dynamic
control flow fits into it.

The core {\tf} runtime is implemented in C++ for portability and performance.
The runtime exports a client API to front-ends for various
languages (such as Python, Go, Java, and C++), whose role is to provide a
high-level interface for building a dataflow graph, and manage the execution
of that graph across a set of workers. In Section~\ref{sec:impl}, we describe 
how the control-flow constructs in the client program are compiled into a
dataflow graph representation for optimization and execution.




The {\tf} runtime is responsible for the execution of the dataflow graph. It
includes optimizations such as common subexpression elimination and constant
propagation. To enable distributed execution on heterogeneous systems, a central
coordinator automatically maps the nodes in the graph to the given set of
devices, then partitions the graph into a set of subgraphs, one per device. When
this partitioning would cut an edge between two devices, it automatically
replaces the edge with a pair of communication operations, \texttt{Send(t, k)}
and \texttt{Recv(k)}, which share a \textit{rendezvous key}. After
partitioning, each subgraph is shipped to the corresponding device, then
executed by a device-local \emph{executor}. The local executors communicate
directly among themselves using \texttt{Send} and \texttt{Recv} operations, with
no involvement from the central coordinator. When a tensor needs to be transported
across devices, \texttt{Send(t, k)} publishes the tensor \texttt{t} under the
rendezvous key \texttt{k} on the sender's device, and \texttt{Recv(k)} pulls the
tensor under key \texttt{k} from the sender's device, blocking until it has been
produced if necessary.

Dynamic control flow introduces additional challenges in the
above design. Without control flow, each operation in the
graph executes exactly once, so every value can be assigned a
unique name, and every \texttt{Send}/\texttt{Recv} pair a
unique rendezvous key at graph-construction time. With control flow, this
property no longer holds: an operation in a loop can execute zero or more times,
and therefore the unique names and rendezvous keys must be generated dynamically
to distinguish multiple invocations of the same operations.

One attractive feature of {\tf} is that it imposes no restrictions on
partitioning: an operation can be assigned to a device provided the
device has the capability to run the corresponding computation,
independently of graph topology or other considerations. By design, our control-flow
extensions preserve this important feature. Therefore, a conditional
branch or a loop body can be arbitrarily partitioned to run across many
devices. For conditional computation, we must provide a mechanism to
inform any waiting \texttt{Recv} operations on untaken branches, to reclaim
resources. Similarly, for iterative computation, we must provide a mechanism
to allow a loop-participating partition to start its next iteration or terminate
its computation. The next section explains how we realize these
behaviors.

TensorFlow does not aspire
to provide fine-grained fault tolerance, and the iterative programs that
use our mechanism typically run to completion between coarse-grained
checkpoints. Because the typical duration
of a computation using our control-flow constructs is much shorter than the
expected mean time between failures, we rely on {\tf}'s coarse-grained
checkpointing mechanism without changes.

%% file: impl.tex
\section{Design and Implementation}\label{sec:impl}

We rely on a small set of flexible,
expressive primitives that serve as a compilation target for
high-level control-flow constructs within a dataflow model of
computation. We explain these primitives in
Section~\ref{sec:impl:prims}.
In Section~\ref{sec:impl:compile}, we then
describe how we use these primitives for compilation.  Finally, in
Sections~\ref{sec:impl:local} and~\ref{sec:impl:dist}, we describe our
design and implementation for local and distributed execution, respectively.

  

\subsection{Control-Flow Primitives}\label{sec:impl:prims}

\begin{figure}
\begin{center}
  \includegraphics[width=\linewidth]{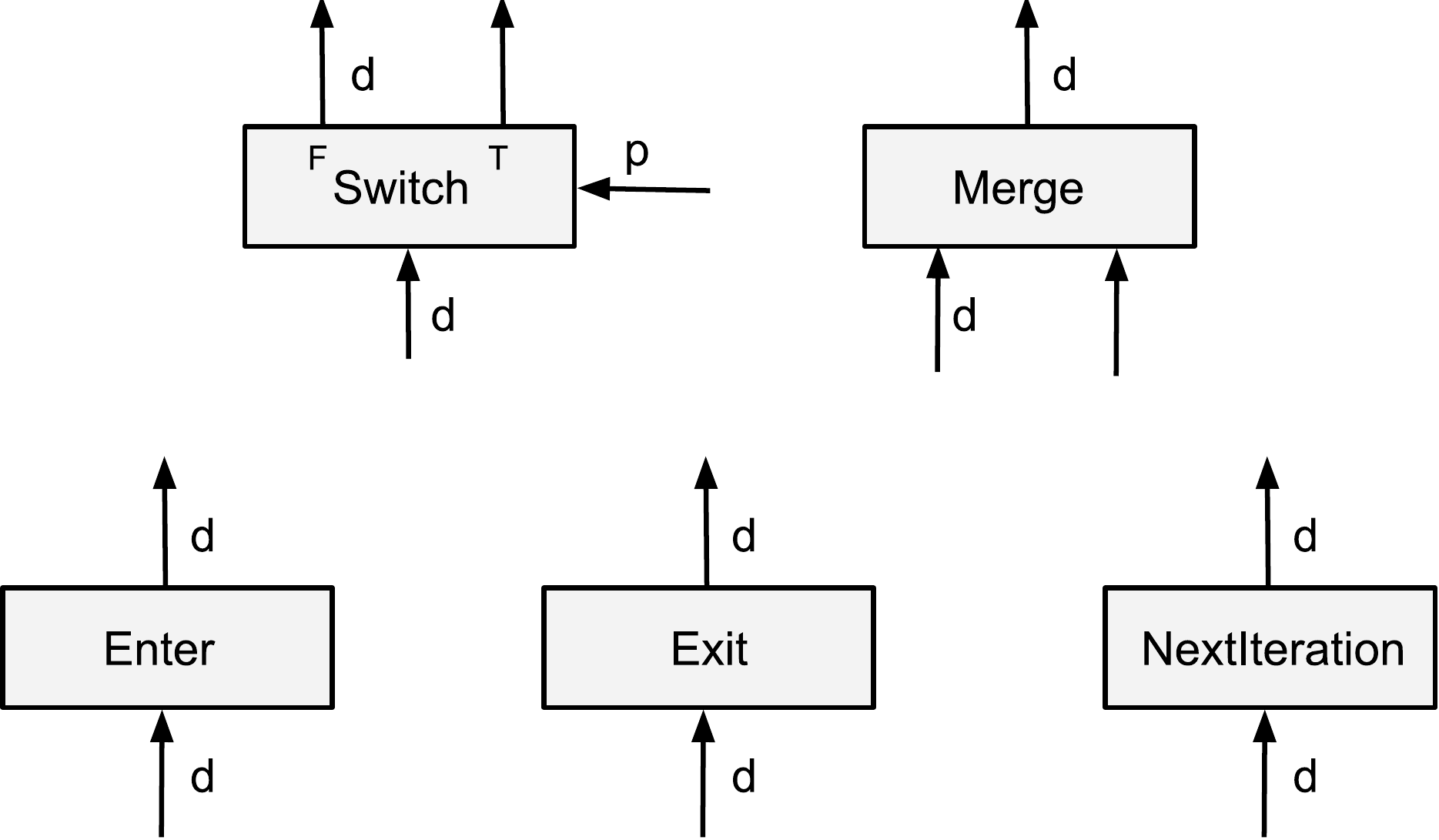}
\end{center}
\caption{The control-flow primitives.}\label{fig:impl:cf_prims}
\end{figure}


Figure~\ref{fig:impl:cf_prims} shows our control-flow primitives
namely \texttt{Switch},
\texttt{Merge}, \texttt{Enter}, \texttt{NextIteration}, and
\texttt{Exit}.  These
primitives resemble those introduced in the classic dynamic dataflow
machines developed by Dennis~\cite{dennis1975preliminary} and
Arvind et al.~\cite{Arvind:1986,Arvind:1990}.
With these primitives, every
execution of an operation takes place within a ``frame'';
concretely, here, frames are dynamically allocated execution contexts associated
with each iteration of a loop. In {\tf} without control flow, each operation in
the graph executes exactly once; when extended with control flow, each operation
executes at most once per frame.

The following is a brief description of the intended semantics.
\begin{itemize}
\item \texttt{Switch} takes a data input \texttt{d} and a boolean
  input \texttt{p}, and forwards the data to one of its outputs \texttt{d\textsubscript{f}} or
  \texttt{d\textsubscript{t}}, based on the value of predicate \texttt{p}.

\item \texttt{Merge} forwards one of its available inputs \texttt{d\textsubscript{1}} or
  \texttt{d\textsubscript{2}} to its
  output; \texttt{Merge} is unlike other TensorFlow primitives in that it is enabled for
  execution when any of its inputs is available.

\item \texttt{Enter} forwards its input to a child 
  frame with a given name. There can be multiple \texttt{Enter} operations
  for the same child frame, each asynchronously making a
  different tensor available to the child frame. The child
  frame is created when the first \texttt{Enter} is executed
  in the runtime.

\item \texttt{Exit} forwards a value computed in a frame to
  its parent frame. There can be multiple \texttt{Exit}s to the
  parent frame, each asynchronously making a tensor available to the
  parent frame.

\item \texttt{NextIteration} forwards its input to the next iteration's
  frame.  Iteration $N+1$ begins when the first \texttt{NextIteration} operation
  is executed at
  iteration~$N$. There can be multiple
  \texttt{NextIteration}s in a frame.
\end{itemize}


\subsection{Compilation of Control-Flow Constructs}\label{sec:impl:compile}

We compile high-level control-flow constructs into dataflow graphs
that comprise the primitives presented above. Next we outline the
basics of the graph construction for conditionals and while-loops.

The translation of \texttt{cond(p, true\_fn, false\_fn)} uses only
\texttt{Switch} and \texttt{Merge}. We invoke \texttt{true\_fn} and
\texttt{false\_fn} respectively to construct the two branches of a
\texttt{cond}. During the construction of a branch, we capture any
external tensor (not created in the branch), and insert a
\texttt{Switch} to guard its entering into the branch. The guard ensures
that any operations in a branch will be executed only when that branch
is taken. The external
tensors may become available at very different times; we use one
\texttt{Switch} for each external tensor in order to maximize parallelism. Each
branch may return several tensors, and both branches must return the
same number and type of tensors. For each output, we add a \texttt{Merge} in
order to merge the true and false branches, thus
enabling downstream computation as soon as possible.

\begin{figure}
\begin{center}
  \includegraphics[width=0.75\linewidth]{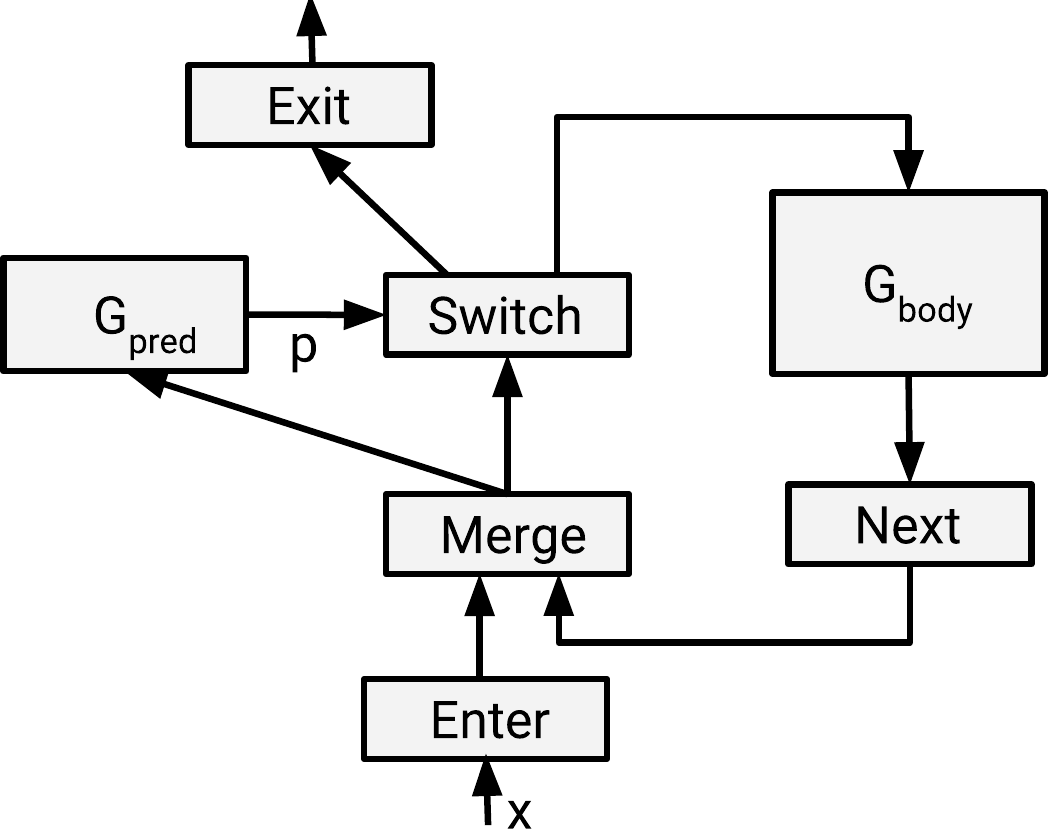}
\end{center}
\caption{Dataflow graph for a while-loop.}\label{fig:impl:while_graph}
\end{figure}

Figure~\ref{fig:impl:while_graph} sketches the dataflow graph for
a while-loop 
with a single loop
variable. (If there were multiple loop variables, there would be a
separate set of \texttt{Enter}, \texttt{Merge}, \texttt{Switch},
\texttt{NextIteration}, and \texttt{Exit} nodes for each of them, so
that multiple iterations can run in parallel.) The loop predicate and the loop body are represented by the subgraphs
$\texttt{G}_{\small   \texttt{pred}}$ and $\texttt{G}_{\small \texttt{body}}$, respectively.
The output of \texttt{Merge} is used as the input to
$\texttt{G}_{\small \texttt{pred}}$ to compute the loop termination condition \texttt{p},
and as an input to \texttt{Switch}, which forwards the tensor either to
\texttt{Exit} to terminate the current loop
or to $\texttt{G}_{\small \texttt{body}}$ to start a new iteration.
The graph
is cyclic so the result of the loop body can go from one iteration to
the next.
Any external tensors used in $\texttt{G}_{\small \texttt{pred}}$ or
$\texttt{G}_{\small \texttt{body}}$ are captured and treated as loop constants. We
automatically insert an \texttt{Enter} for each external tensor to
make it accessible in the loop body.
When a new iteration is started, all the loop constants become available
to that iteration automatically.


This approach supports arbitrary nestings of conditionals and loops. For
example, a loop body can contain a conditional.

\subsection{Local Execution}\label{sec:impl:local}


Recall that {\tf} partitions a dataflow graph into a set of
subgraphs. Each subgraph runs on a separate device, managed by a
local executor that runs on the device's host. (The host's CPU is also
represented as a device.)
This subsection describes how such a
local executor can support dynamic control flow.

The local executor is conceptually straightforward. It starts from the
source nodes and repeatedly executes the nodes that become ready.  A
node, with the exception of \texttt{Merge}, becomes ready when all its
inputs are available. All \texttt{Recv} nodes in a subgraph
are regarded as source nodes.

For graphs without control-flow constructs, every node is executed exactly once
and the execution ends when all nodes have been executed. Dynamic control flow
introduces new complexity. An operation now can be executed any
number of times. The executor must manage the (possibly
concurrent) execution of multiple instances of the same operation, and to
determine the completion of the entire execution.

We redesigned the local executor of {\tf} to handle dynamic control flow. In
order to track the tensors generated by different invocations of the same
operation, each tensor inside the executor is represented as a tuple
$(\textit{value}, \textit{is}\_\textit{dead}, \textit{tag})$, where
$\textit{value}$ is a tensor value, $\textit{is}\_\textit{dead}$ is a boolean
that indicates whether the tensor is on an untaken branch of a \texttt{Switch},
and $\textit{tag}$ is a globally unique identifier for the tensor. Intuitively,
the tag defines a dynamic execution context---a frame, in the terminology of
Section~\ref{sec:impl:prims}. Each loop iteration starts a new frame, and within
a frame an operation is executed at most once. As a result, the tag is used to
distinguish the tensors generated by different iterations. This distinction is
critical for the correct rendezvous of \texttt{Send} and \texttt{Recv}
operations, since tags are used as the rendezvous keys.

\begin{figure}
\begin{verbatim}
Eval(Switch(p, d), c) = (r1, r2), where
  r1 = (value(d), p || is_dead(d), tag(d))
  r2 = (value(d), !p || is_dead(d), tag(d))

Eval(Merge(d1, d2), c) = r, where
  r = if is_dead(d1) then d2 else d1

Eval(Enter(d, name), c) = r, where
  r = (value(d), is_dead(d), tag(d)/name/0)

Eval(Exit(d), c) = r, where
  r = (value(d), is_dead(d), c.parent.tag)

Eval(NextIteration(d), c) = r, where
  tag(d) = tag1/name/n
  r = (value(d), is_dead(d), tag1/name/(n+1))

Eval(Op(d1, ..., dm), c) = (r1, ..., rn), where
  value(r1), ..., value(rn) =
    Op(value(d1), ..., value(dm))
  is_dead(ri) =
    is_dead(d1) || ... || is_dead(dm), for all i
  tag(ri) = tag(d1), for all i
\end{verbatim}
\caption{Evaluation rules for control-flow operators.}\label{fig:impl:eval_rules}
\end{figure}

The executor implements the evaluation rules shown in
Figure~\ref{fig:impl:eval_rules}. Each evaluation rule
\texttt{Eval(e, c) = r} describes how to evaluate expression \texttt{e} in
frame \texttt{c}, yielding output \texttt{r}.
The operations \texttt{Enter} and \texttt{Exit} create and delete execution frames, respectively;
for simplicity, the rules do not show this.
All the inputs to
an operation must have the same matching tag; \texttt{c.parent}
is \texttt{c}'s parent frame, with \texttt{c.parent.tag} as its tag. 

The last rule applies to all non-control-flow operations. In the implementation,
the actual computation is performed only when none of the inputs are
dead. If there is a dead input, we skip the computation and
propagate a dead signal downstream. This propagation of deadness is
useful for supporting distributed execution, as explained in
the next subsection. The choice to propagate the tag of the first input is
arbitrary; all inputs must have the same tag.

While these rules allow multiple loop iterations to run in
parallel, more parallelism typically results in
more memory consumption. We therefore incorporate knobs in the local
executor that
allow us to limit the degree of parallelism. In our evaluation, we generally
find that a limit of 32 works well---better than 1, which would be easy to
achieve with stricter implementation strategies, but also better than no limit
at all. The optimal value depends on the details of the hardware and the model.



\subsection{Distributed Execution}\label{sec:impl:dist}

The main challenge for distributed, dynamic control flow arises when
the subgraph of a conditional branch or loop body is partitioned
across devices. We favor a design that allows the executors of
the partitions to make progress independently, with no centralized
coordinator. We do not require synchronization after
loop iterations, as this would severely limit parallelism.  Each device
that participates in a loop can start the next iteration or exit once
it receives the value of the loop predicate. A partition can have multiple
iterations running in parallel, and partitions can work on different
iterations of the same loop.
The local executors communicate only via \texttt{Send} and
\texttt{Recv} operations.  A centralized coordinator is involved only in
the event of completion or failure.

We first look at conditionals, and consider a \texttt{Recv} operation on an
untaken branch. A \texttt{Recv} is always ready and can be started
unconditionally. So the system would block, without reclaiming resources, if the
corresponding \texttt{Send} is never executed. The solution is to propagate an
\texttt{is\_dead} signal across devices from \texttt{Send} to \texttt{Recv}. The
propagation may continue on any number of devices. This propagation
scheme handles distributed execution of nested conditionals, and interacts well
with distributed execution of loops. However, when there are many
\texttt{Send}--\texttt{Recv} pairs across devices on a rarely taken
conditional branch, the large number of \texttt{is\_dead} signals may cause
performance issues. These situations seldom arise, but we have
prototyped an optimization that transmits an \texttt{is\_dead} signal only once
to the destination device and, there, broadcasts it to all \texttt{Recv}
operations.

\begin{figure}
\begin{center}
  \includegraphics[width=\linewidth]{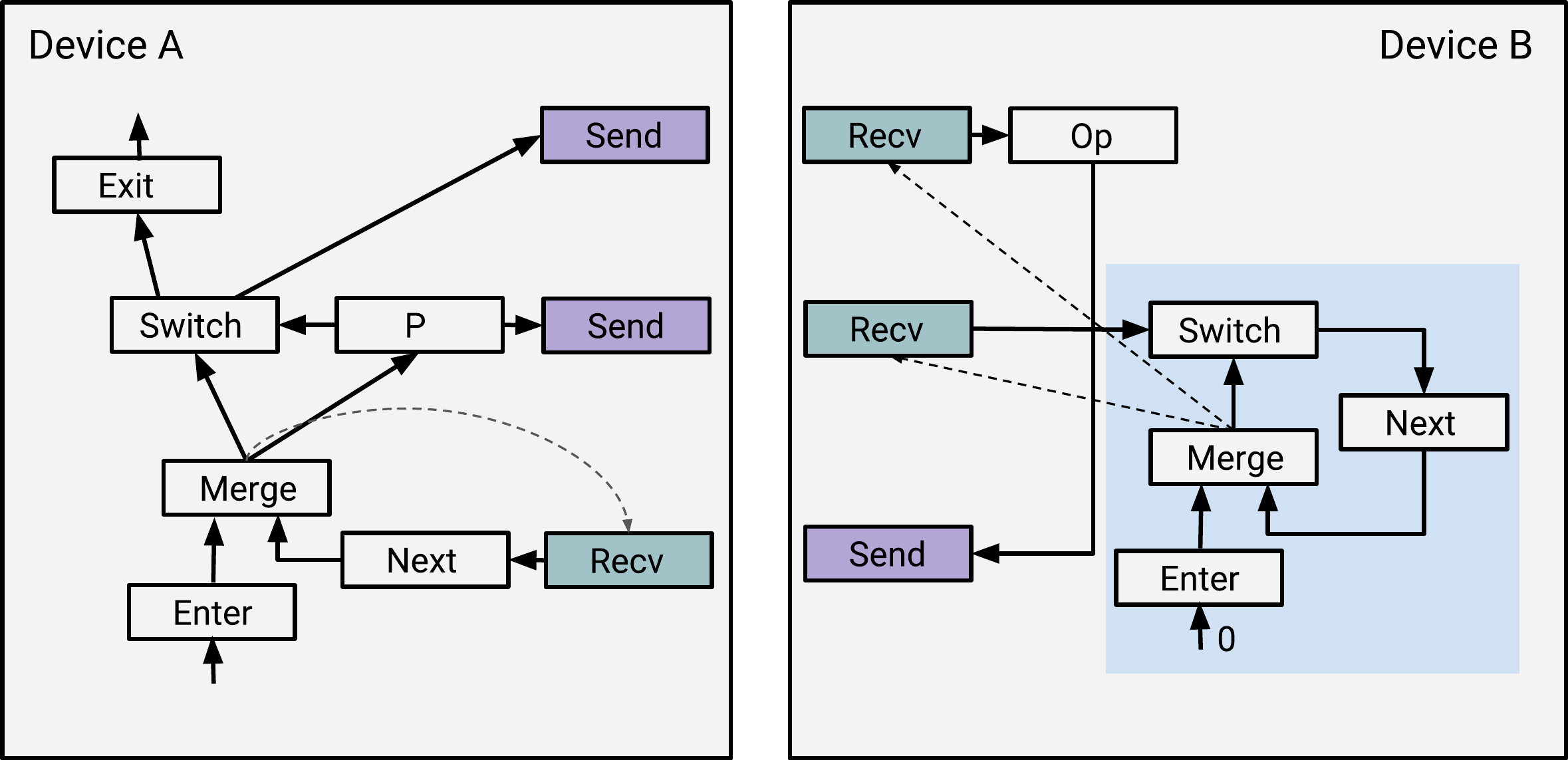}
\end{center}
\caption{Distributed execution of a while-loop.}\label{fig:impl:while_partition}
\end{figure}

For the distributed execution of a loop, at each iteration, each partition
needs to know whether to proceed or exit. We address this need by automatically
rewriting the graph with simple control-loop state machines.
%
%
%
%
For example,
Figure~\ref{fig:impl:while_partition} shows the result of
partitioning a simple while-loop across two devices. 
The loop body contains only one operation \texttt{Op} assigned
to device~B. In B's partition we add a control-loop state machine (blue shade in the figure),
which controls the \texttt{Recv} operations inside
the loop. The dotted lines are control edges, and impose an order on operations.
Suppose that device A is executing the loop predicate
\texttt{P} at iteration~$i$;  distributed execution may then proceed as follows:
\begin{itemize}
\item On A,  \texttt{Recv} awaits a
  value from B. A sends \texttt{P} to B, so B knows the
  decision on iteration $i$ and, depending on whether \texttt{P} is true,
  A sends either the input tensor for \texttt{Op} or a dead signal to~B. 
  
\item On B, if \texttt{Recv} for \texttt{Op} gets a real
  tensor from  A, B executes \texttt{Op} and sends back a real
  tensor. Otherwise, if the \texttt{Recv} gets an \texttt{is\_dead} signal, B
  propagates the signal through \texttt{Op} and sends an \texttt{is\_dead}
  signal back to A.
  If \texttt{Recv} for \texttt{Switch} gets true, the
  control-loop state machine  further enables \texttt{Recv}s
  for the next iteration. Otherwise, the control loop terminates,
  and so does execution on B.

\item Back on A, if \texttt{Recv} gets a real tensor,
  the next iteration is started. Otherwise, execution terminates.
\end{itemize}

The overhead for the distributed
execution of a loop is that every participating device needs to
receive a boolean at each iteration from the device that produces the
loop predicate. However, the communication is asynchronous and computation of the
loop predicate can often run ahead of the rest of the computation.
Given typical neural network models, this overhead is minimal and largely
hidden.

To make things concrete, let us take a brief look at GPU execution.
In this setting, the compute and I/O operations typically run on GPU as a sequence of
asynchronous kernels, whereas control-flow decisions are made
by the local executor for the GPU on the host.  From the point of view of this local executor, a GPU
kernel is considered completed once it is enqueued into the correct
GPU stream. (Correctness is guaranteed by the sequential execution
of the kernels in a single stream and proper cross-stream
synchronization.) So once we allow parallel iterations, the local executor
will typically run completely in parallel to the compute and I/O
operations and on separate computing resources. Therefore, as we will show in
Section~\ref{sec:eval}, dynamic control flow gives the same performance
as static unrolling.

%
%
%
%
%
%

%% file: autodiff.tex
\section{Automatic Differentiation and Memory Management}\label{sec:diff}

Machine learning algorithms often rely on gradient-based methods
for optimizing a set of parameters. During the training of models, gradient computations
usually take more than half of the compute time.
It is therefore critical to make these computations efficient and scalable. 

{\tf} supports automatic differentiation: given a dataflow graph that
represents a neural network, {\tf} generates efficient code for the
corresponding distributed gradient computations. This section
describes how we extend automatic differentiation to control-flow
constructs, and (briefly) the treatment of
TensorArrays.

This section also describes techniques for memory management.
Although these techniques are motivated by automatic
differentiation, they are not specific to this purpose.



\subsection{Backpropagation with Control Flow}\label{sec:diff:backprop}

{\tf} includes a reverse-mode automatic differentiation (autodiff) library that
implements the well-known backpropagation
algorithm~\cite{rumelhart1988learning}, and we describe here its
support for (potentially nested) \texttt{cond} and \texttt{while\_loop}
constructs.

We first revisit the basic autodiff algorithm used in
{\tf}. The \texttt{tf.gradients()} function computes the gradients of a scalar
function, $f(x_1, x_2, \ldots)$, with respect to a set of parameter tensors,
$x_1, x_2, \ldots$, using the following algorithm, which implements the
vector chain rule:\footnote{Parr and Howard survey the necessary calculus to
understand this algorithm~\cite{parr2018calculus}.}

\begin{figure}
\begin{center}
  \includegraphics[width=\linewidth]{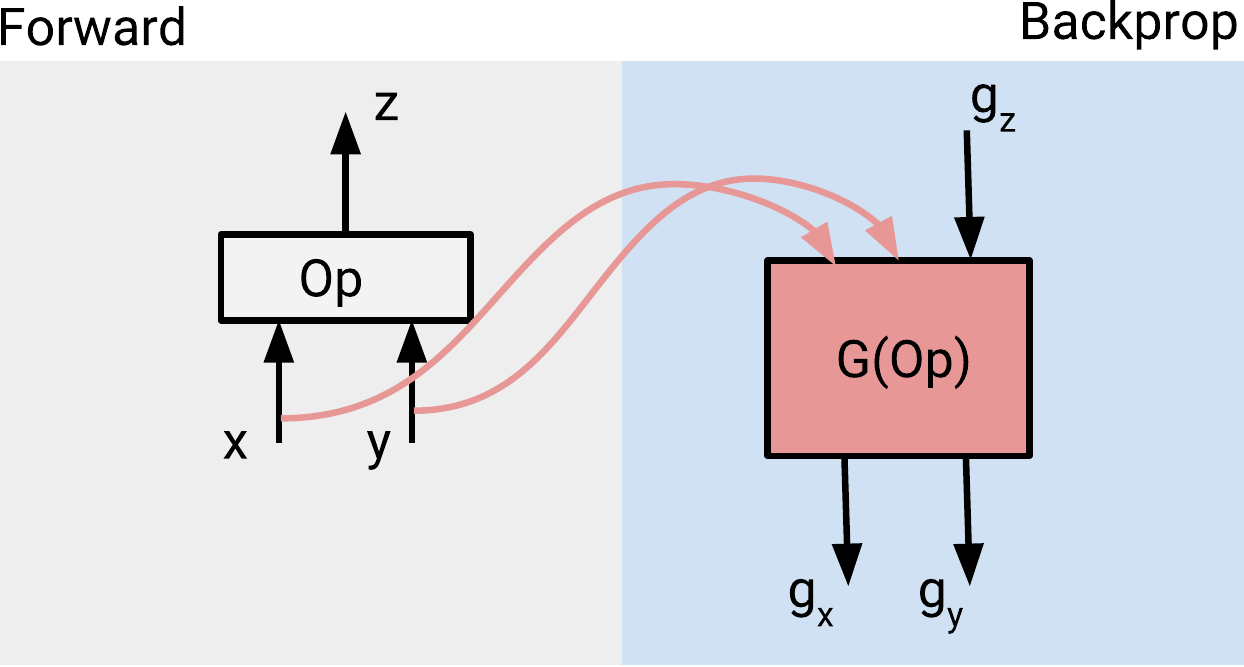}
\end{center}
\caption{An operation and its gradient function.}\label{fig:diff:op_grad}
\end{figure}

\begin{enumerate}
\item Identify the subgraph $G$ of operations between the symbolic tensor
  representing $y = f(x_1, x_2, \ldots)$ and each parameter tensor
  $x_1, x_2, \ldots$.
\item For each edge in $G$, which represents an intermediate value $t$ in $f$,
  set $\mathrm{Grads}[t] := 0$. Set $\mathrm{Grads}[y] := 1$.
\item Traverse the vertices of $G$ in reverse topological order. For each vertex
  representing an intermediate operation
  \[
  o_1, \ldots, o_n = \mathrm{Op}(i_1, \ldots, i_m),
  \] invoke the corresponding ``gradient function''
  \[
  \partial
  i_1, \ldots, \partial i_m = \mathrm{OpGrad}(i_1, \ldots, i_m,
  \mathrm{Grads}[o_1], \ldots, \mathrm{Grads}[o_n]).
  \] Add each $\partial i_k$ to
  $\mathrm{Grads}[i_k]$.
\item After traversing every vertex, the gradient
  $\frac{\partial y}{\partial x_k}$ is $\mathrm{Grads}[x_k]$.
\end{enumerate}

{\tf} includes a library of gradient functions that correspond to most of its
primitive operations. Figure~\ref{fig:diff:op_grad} shows a schematic example
of such a gradient function, \texttt{G(Op)}, which depends on both the partial
derivative of $y$ with respect to the original \texttt{Op}'s output
($\texttt{g}_{\small \texttt{z}}$) and the original \texttt{Op}'s inputs
(\texttt{x} and \texttt{y}). As a concrete example, the \texttt{tf.matmul(x, y)}
operation has the following (simplified) gradient function, which computes
gradients with respect to matrices \texttt{x} and \texttt{y}:

\begin{verbatim}
# NOTE: `op` provides access to the inputs
# and outputs of the original operation.
def MatMulGrad(op, g_z):
  x, y = op.inputs
  g_x = tf.matmul(g_z, tf.transpose(y))
  g_y = tf.matmul(tf.transpose(x), g_z)
  return g_x, g_y
\end{verbatim}

\noindent Tensors \texttt{x} and \texttt{y} are used in the gradient function,
so will be kept in memory until the gradient computation is performed.
The resulting memory consumption is a major constraint on our ability to
train deep neural networks, as we need values from all layers of the forward
computation (that is, of the computation being differentiated)
for the gradient computation. The same difficulty applies to neural networks
with loops. We return to this topic in Section~\ref{sec:diff:memory}.


We describe here the mechanisms that enable {\tf}'s autodiff algorithm to perform backpropagation through
control-flow constructs. Each operation in the graph is associated
with a ``control-flow context'' that identifies the innermost control-flow
construct of which that operation is a member. When the backpropagation
traversal first encounters a new control-flow context, it generates a
corresponding control-flow construct in the gradient graph.

The gradient for \texttt{tf.cond(pred, true\_fn, false\_fn)} with
output gradients \texttt{g\_z} is
\[
\texttt{tf.cond(pred, true\_fn\_grad(g\_z),
false\_fn\_grad(g\_z))}.
\]
To obtain the \texttt{true\_fn\_grad} function, we
apply \texttt{tf.gradients()} to the symbolic outputs $t_i$ of
\texttt{true\_fn}, but setting the initial gradients $\mathrm{Grads}[t_i] :=
\texttt{g\_z}[i]$; the same logic applies to \texttt{false\_fn\_grad}.

The logic for \texttt{tf.while\_loop(cond, body, loop\_vars)} is more
complicated. To understand its key components, consider the following simple
program that iteratively multiplies an input matrix \texttt{x} by a parameter
matrix \texttt{w}:

\begin{verbatim}
w = tf.Variable(tf.random_uniform((10, 10))
x = tf.placeholder(tf.float32, (10, 10))
a = tf.while_loop(
    lambda i, a_i: i < 3,
    lambda i, a_i: (i + 1, tf.matmul(a_i, w)),
    [0, x])
y = tf.reduce_sum(a)
g_w = tf.gradients(y, x)
\end{verbatim}

\begin{figure}
\begin{verbatim}
w = tf.Variable(tf.random_uniform((10, 10))
x = tf.placeholder(tf.float32, (10, 10))
a_0 = x
a_1 = tf.matmul(a_0, w)
a_2 = tf.matmul(a_1, w)
a_3 = tf.matmul(a_2, w)
y = tf.reduce_sum(a_3)
g_y = 1.0
g_w = 0.0
# The gradient of reduce_sum is broadcast.
g_a_3 = tf.fill(tf.shape(a_3), g_y)
# Apply MatMulGrad() three times; accumulate g_w.
g_a_2 = tf.matmul(g_a_3, tf.transpose(w))
g_w += tf.matmul(tf.transpose(a_2), g_a_3)
g_a_1 = tf.matmul(g_a_2, tf.transpose(w))
g_w += tf.matmul(tf.transpose(a_1), g_a_2)
g_a_0 = tf.matmul(g_a_1, tf.transpose(w))
g_w += tf.matmul(tf.transpose(a_0), g_a_1)
\end{verbatim}
\caption{Computing the gradient of a loop by unrolling.}
\label{fig:autodiff:unrolled}
\end{figure}

In this simple example, we can take advantage of the loop bound being
a constant. Figure~\ref{fig:autodiff:unrolled} illustrates how we might unroll
the loop statically and
apply the gradient functions for
\texttt{tf.matmul()} and \texttt{tf.reduce\_sum()}:\footnote{In the general case, the loop bound
  may depend on the input data---e.g., based on the length of a
  sequence in an RNN---and we must construct a
  \texttt{tf.while\_loop()} for the gradients.}

This example highlights three features of the general solution:

\begin{enumerate}
\item The gradient of a \texttt{tf.while\_loop()} is another loop that executes
  the gradient of the loop body for the same number of iterations as the forward
  loop, but in reverse. In general, this number is not known statically, so
  the forward loop must be augmented with a loop counter.
\item The gradient of each differentiable loop variable becomes a loop variable
  in the gradient loop. Its initial value is the gradient with respect to the
  corresponding loop output (e.g., \texttt{g\_a\_3} in the example).
\item The gradient of each differentiable tensor that is constant in the loop
  (e.g., \texttt{g\_w} in the example) is the sum of the gradients for that
  tensor at each iteration.
\end{enumerate}

\begin{figure}
\begin{center}
  \includegraphics[width=\linewidth]{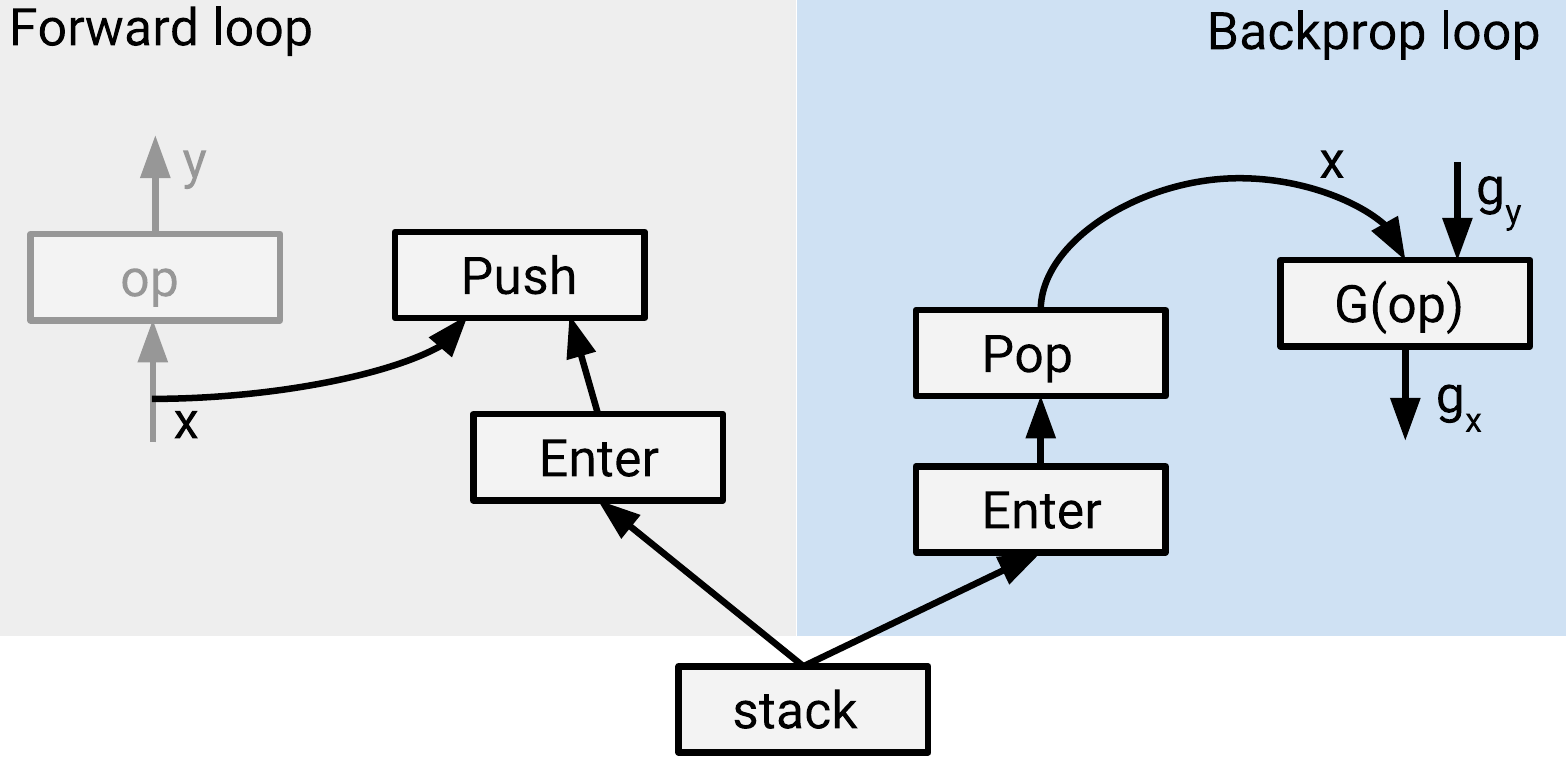}
\end{center}
\caption{Saving a tensor for reuse in backpropagation.}\label{fig:diff:stack}
\end{figure}

\noindent
In addition, intermediate values \texttt{a\_1}, \texttt{a\_2}, and
\texttt{a\_3} from the forward loop are used in the gradient loop.
The performance of our solution depends heavily on how we treat such
intermediate values. There are typically many such values, such as the inputs of
a matrix multiplication or the predicate of a conditional nested in the
loop. We avoid the computational expense of recomputing these values by
automatically rewriting the forward loop to save any intermediate values that
the gradient loop needs.
We introduced a stack data structure into {\tf} to save values across loops:
the forward computation pushes onto the stacks, the gradient computation pops.
Figure~\ref{fig:diff:stack} shows the graph structure that implements this
stack-based state saving. The implementation uses a different stack for each
intermediate value that is reused in the gradient loop, in order to allow
individual gradients to be computed asynchronously.
We prefer to use a regular tensor to store each intermediate value,
because each intermediate value might have a different shape, and packing the
values into a dense, contiguous array would incur unnecessary memory copies.
However, if the loop variables have a static shape and the iteration
count has a static upper bound, the XLA~\cite{xla} compiler may lower the stack
operations to read/write operations on a contiguous mutable array.


Correctness requires care to preserve the proper ordering of stack operations.
Performance considerations lead us to making
the stack operations asynchronous, so that they can run in parallel
with the actual computation.
For example, in Figure~\ref{fig:diff:stack}, \texttt{Op} (and even
operations in subsequent iterations) can potentially run in parallel with
\texttt{Push}. As Section~\ref{sec:diff:memory} explains, this asynchrony
is important for overlapping compute and I/O operations. For correctness, we add
explicit control dependencies to enforce ordering of the stack operations.

Our implementation has additional optimizations, often with the goal of reducing
memory usage. For example, if a value is immediately reduced in the gradient
code by an operation that computes its shape, rank, or size, we move this
operation into the forward loop. Moreover, for calculations that accumulate
gradients, we introduce subgraphs that sum gradients eagerly into new loop
variables.


Our approach also accommodates nested control-flow constructs. When a
conditional nests inside a while-loop, we push the guard values at all
forward iterations onto a stack, and pop those values to control the
conditionals in the gradient loop. For nested loops, we apply our techniques
recursively.
Although we hope this explanation gives the
intuition for control-flow operator gradients, a rigorous
mathematical treatment is beyond the scope of this paper. The literature on
automatic differentiation has considered the question of correctness of the
semantics of control flow with respect to the mathematical notion of
differentiation (e.g.,~\cite{Beck:1994:IAD:195968.196017}), and our algorithms
follow these established principles. That said, this question remains a subject
of ongoing research~\cite{gordtube}.

\subsection{Backpropagation with TensorArrays}\label{sec:diff:tensorarray}

TensorArrays constitute an important element of our programming model,
so automatic differentiation must treat them correctly and efficiently.
For this purpose, we require that
each location of a TensorArray may be written only once in the forward computation being differentiated,
but allow multiple reads from the same location. This requirement is satisfied by
common applications of TensorArrays.

The {\tf} runtime represents TensorArrays as ``resource objects'', which are containers for
mutable state. Each TensorArray \texttt{ta} exposes an opaque handle \texttt{ta.handle}. Operations such as \texttt{write} and \texttt{read}
accept a TensorArray handle as their primary argument.
During backpropagation, for each forward TensorArray a new
TensorArray of the same size is created to hold
gradient values.  The operation \texttt{ta.grad()}, when
executed, either creates or performs a table lookup for the
gradient TensorArray associated with handle \texttt{ta.handle}.
The TensorArray operations are duals of each other:
the gradient of \texttt{ta.read(ix)} is
\texttt{ta.grad().write(ix, $\texttt{g}_{\small \texttt{ix}}$)}, and vice versa, 
and the gradient of \texttt{ta.unstack(ts)} is \texttt{ta.grad().stack()}, and vice versa.
When there are multiple reads to the same location, the gradient TensorArray holds the sum of the partial gradients generated by the reads.
Our implementation ensures the proper ordering
of reads and writes while allowing parallelism.

\begin{figure*}[t]
\centering
  \begin{tabular}[b]{c}
    \includegraphics[width=.29\linewidth]{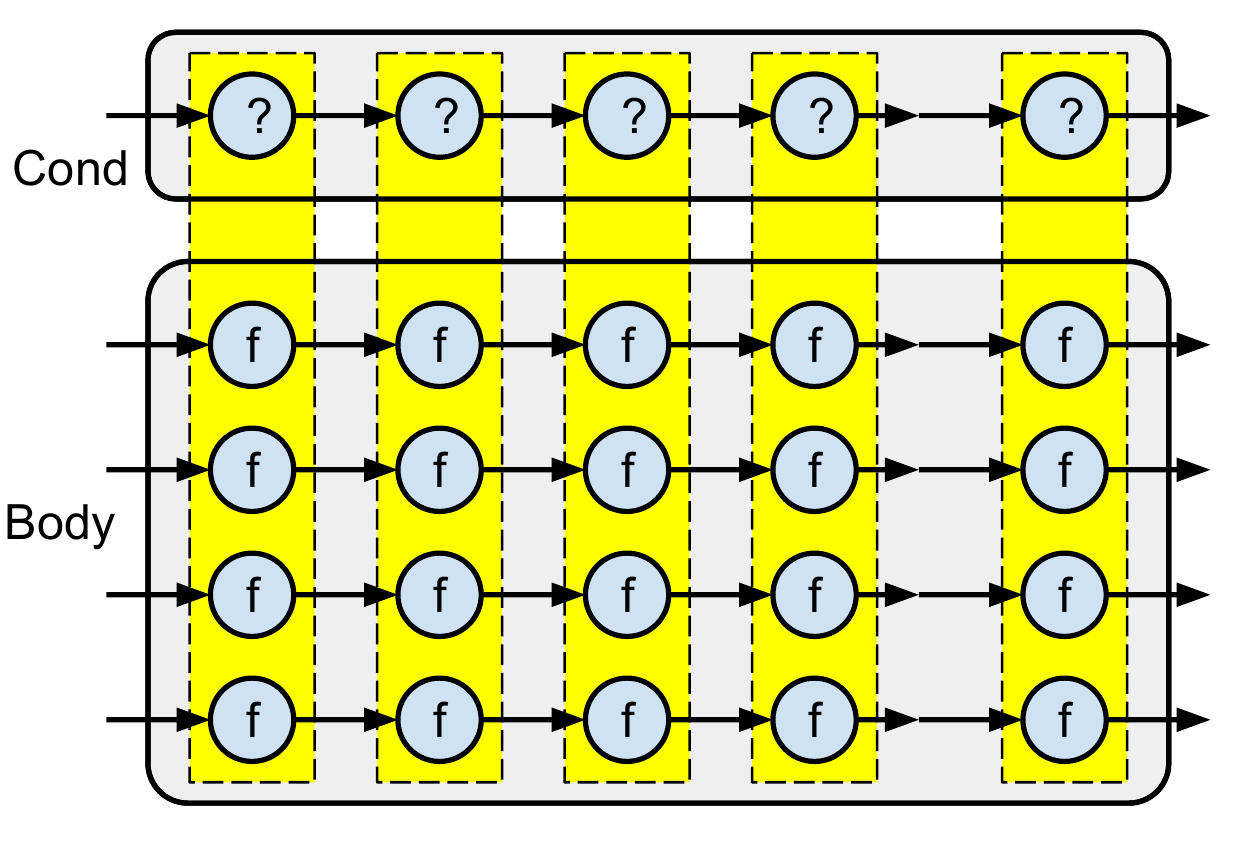}\\
    \small (a) Independent devices
  \end{tabular}
  \begin{tabular}[b]{c}
    \includegraphics[width=.32\linewidth]{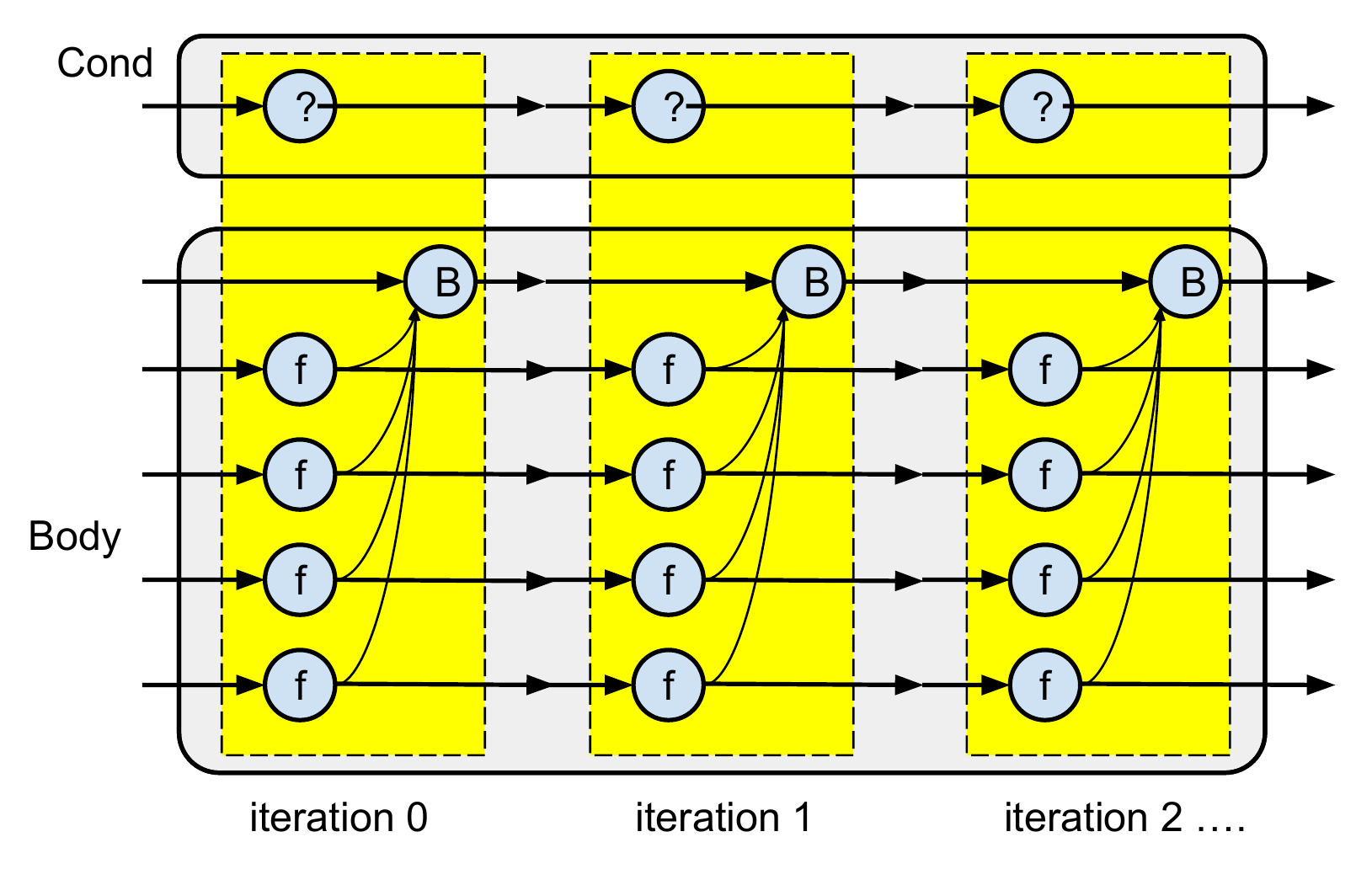}\\
    \small (b) Barrier / AllReduce
  \end{tabular}
  \begin{tabular}[b]{c}
    \includegraphics[width=.32\linewidth]{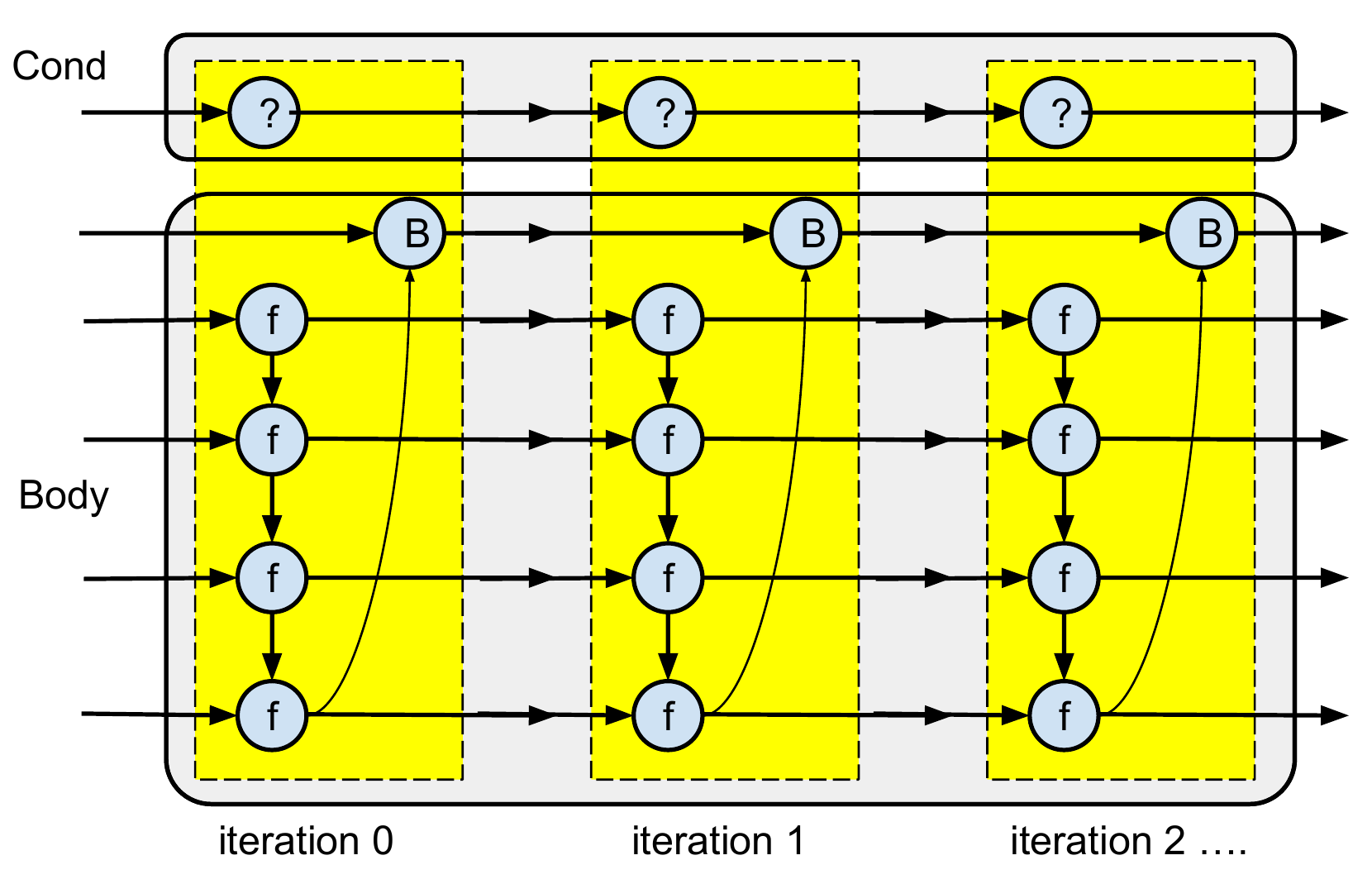}\\
    \small (c) Data-dependent loop body
  \end{tabular}
  \caption{Dataflow dependencies in a distributed while-loop.}\label{fig:eval:data_dependencies}
\end{figure*}

\subsection{Memory Management}\label{sec:diff:memory}

The memory demands described above are critical on specialized devices such as
GPUs, where memory is typically limited to no more than 16GB. We employ several
techniques for alleviating this scarcity. We rely on memory swapping, taking
advantage of temporal locality. We move tensors from GPU to CPU memory (which is
relatively abundant) when they are pushed onto stacks, and bring them back
gradually before they are needed in backpropagation. The key to achieving good
performance in memory swapping is to overlap compute and I/O operations. This
goal requires the seamless cooperation of several system components.

First, as explained above, multiple iterations of a loop can run in parallel,
and the stack push and pop operations are asynchronous and can run in parallel
with the computation proper. The forward computation can run ahead without
waiting for the completion of I/O operations. Conversely, during the gradient
computation, the I/O operations can run ahead, prefetching the tensors that will
be needed next.

Second, in the context of GPUs, we use separate GPU streams for compute
and I/O operations to further improve the overlap of these two
classes of operations. Each stream consists of a sequence of GPU kernels that
are executed sequentially. Kernels on different streams can run in
parallel with respect to each other. We therefore use (at least) three
streams for compute, CPU-to-GPU transfer, and GPU-to-CPU transfer
operations, respectively. Thus, we run compute kernels in
parallel with independent memory-transfer kernels. Of course, special
care must be taken when there is a causal dependency between two
kernels on different streams. We rely on a combination of {\tf} control
edges and GPU hardware events to synchronize the dependent operations
executed on different streams.

Our implementation of the scheme described above watches
the memory consumption reported by the {\tf} memory allocator, and
only starts to swap when memory consumption reaches a predefined
threshold. We also do not swap small tensors or the same value more than once.

This scheme has a large impact.
For instance, as Section~\ref{sec:eval} describes, for an example RNN model for processing sequences, without memory swapping,
we run out of memory for sequences of length
$500$.  With memory swapping, we can handle
sequences of length $1000$ with little overhead, and available host
memory is the primary factor limiting the maximum sequence length.

%% file: eval.tex
\section{Evaluation}\label{sec:eval}


In this section, we evaluate our design and implementation, focusing on key
design choices, and comparing against performance baselines. In particular, our
main comparisons against TensorFlow without in-graph dynamic control flow
are Table~\ref{tbl:eval:swap} (comparing with the system with swapping
disabled), Figure~\ref{fig:eval:static} (comparing with static unrolling), and
Section~\ref{ss:eval:rl} (comparing with out-of-graph control flow). Other
experiments (in particular
Figures~\ref{fig:eval:parallel_iterations} and \ref{fig:eval:model_parallel})
give evidence of the performance benefits of our approach relative to more
simplistic approaches that would limit parallelism or distribution. We focus on
these baselines because they permit the most direct, apples-to-apples
comparison.

For all the experiments, unless otherwise stated, we run {\tf} on a shared
production cluster, consisting of Intel servers with NVidia Tesla K40 GPUs
connected by Ethernet across a production networking fabric, and reported
performance numbers are averages across five or more repeated runs.


\subsection{Data Dependencies}\label{ss:eval:micro}


In this experiment we use simple microbenchmarks to evaluate the performance and
scalability of distributed iterative computation. The benchmark is a single
while-loop with its loop body partitioned to run on a cluster of GPUs. First we
consider two common patterns of data dependence for the loop body: one where
there is no coordination between devices, and one where devices
synchronize at the end of each iteration using a barrier (e.g.,~$AllReduce$), as
illustrated in Figures~\ref{fig:eval:data_dependencies}(a) and
~\ref{fig:eval:data_dependencies}(b).  Such computation patterns are quite common
in multi-layer architectures with MoEs and RNNs.

We evaluate the overall system capacity in terms of the number of iterations per
second it can support when running a while-loop distributed across a set of
GPUs. Each GPU is hosted on a separate machine in the cluster. The computation
$f$ on each GPU is a very small matrix operation, optionally followed by a
barrier $B$ across all GPUs, so this experiment gives us the maximum number of
distributed iterations the system can handle at various scales.

\begin{figure}[t]
  \begin{center}
    \includegraphics[width=\linewidth]{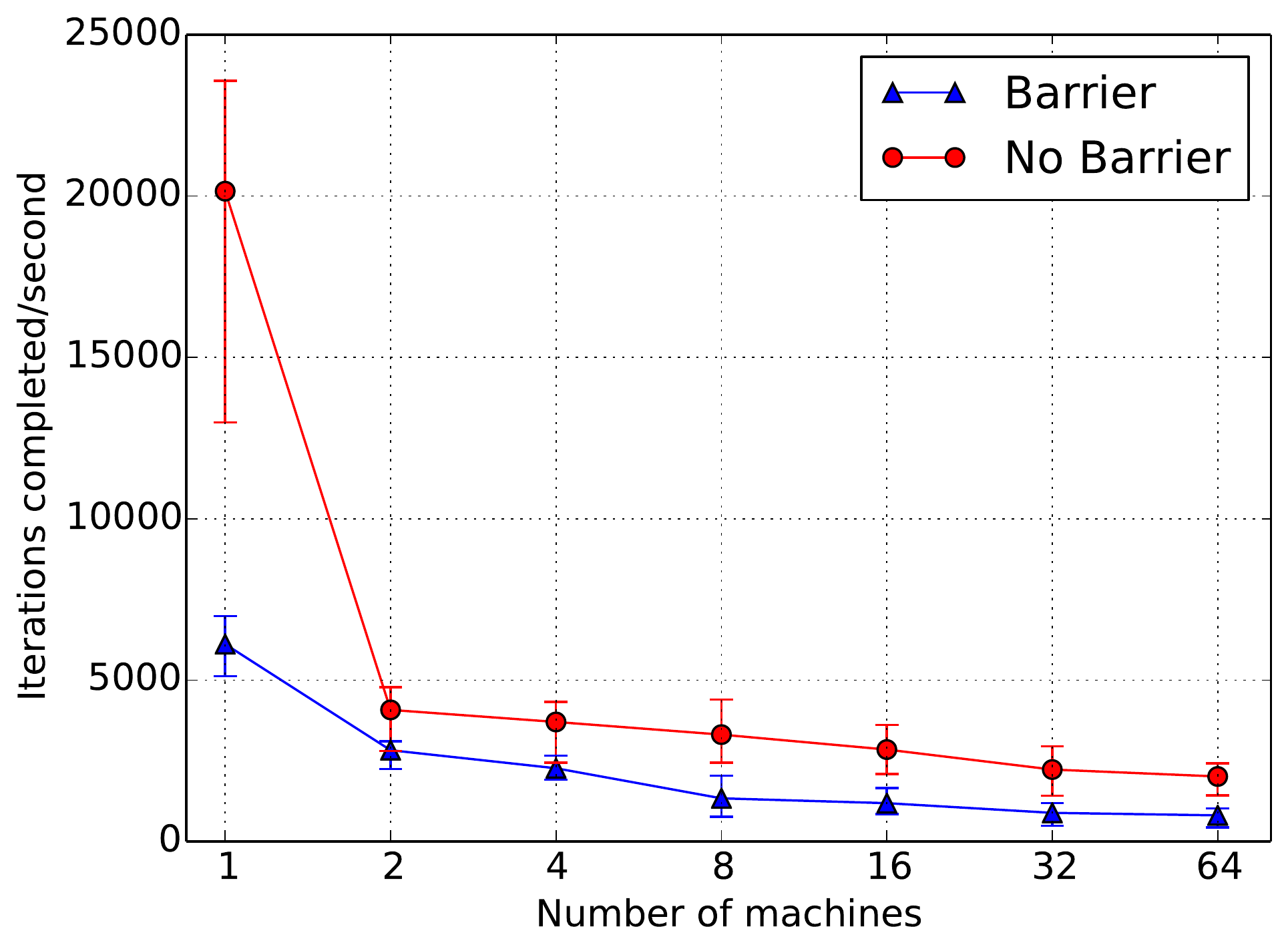}
  \end{center}
  \caption{Performance of a distributed while-loop with a trivial body on a GPU cluster.}\label{fig:eval:distributed_control_flow}
  \vspace{-2mm}
\end{figure}

Figure~\ref{fig:eval:distributed_control_flow} shows the number of iterations
achieved per second as we vary the number of machines from $1$ to $64$. We plot
the median and 5th/95th percentile performance from 5000 trials.  When the loop body
has no cross-device dependencies, the system can support over $20,000$
iterations per second on a single machine, decreasing to $2014$ with $64$
machines. (i.e., $457{\mu}s$ per iteration).  If the loop contains a barrier
operation, this reduces to $809$ iterations per second ($1235{\mu}s$ per
iteration).

Figure~\ref{fig:eval:distributed_control_flow} demonstrates that, for both
patterns of data dependency, the overhead of distributed execution remains
acceptable as the number of machines increases.  The \emph{Barrier} result is
within a factor of two of the global barrier scaling results of Naiad
\cite[Figure~6(b)]{murray2013naiad}, although the variance here is much lower.
Other experiments (in particular, those in Figure~\ref{fig:eval:model_parallel})
further characterize the scaling that results from distributed execution using
non-synthetic workloads.



Next we evaluate the benefits of running multiple iterations in parallel. We run
our simple benchmark on $8$ GPUs in a single server and vary the number
of parallel iterations from $1$ to $32$. The loop consists of $8$ layers of
computation, one for each GPU; and each GPU performs a $1024x1024$ matrix
multiplication before passing the result of its computation to the next.  Each
GPU has a data dependency on its state from the previous loop iteration, and on
the output of the previous GPU. The loop iterations are additionally serialized
as shown in Figure~\ref{fig:eval:data_dependencies}(c).  Note that the loop
condition has no data dependency on the body---when computing on GPUs, this
independence can often allow CUDA kernels from many iterations to be enqueued on
a stream for future execution.

We measure performance both on a K40 equipped server, as in the previous
experiment, and on NVidia's flagship DGX-1 machine equipped with 8 V100 GPUs,
plotting the median and 5th/90th percentiles from 5000 trials.

\begin{figure}[t]
  \begin{center}
    \includegraphics[width=\linewidth]{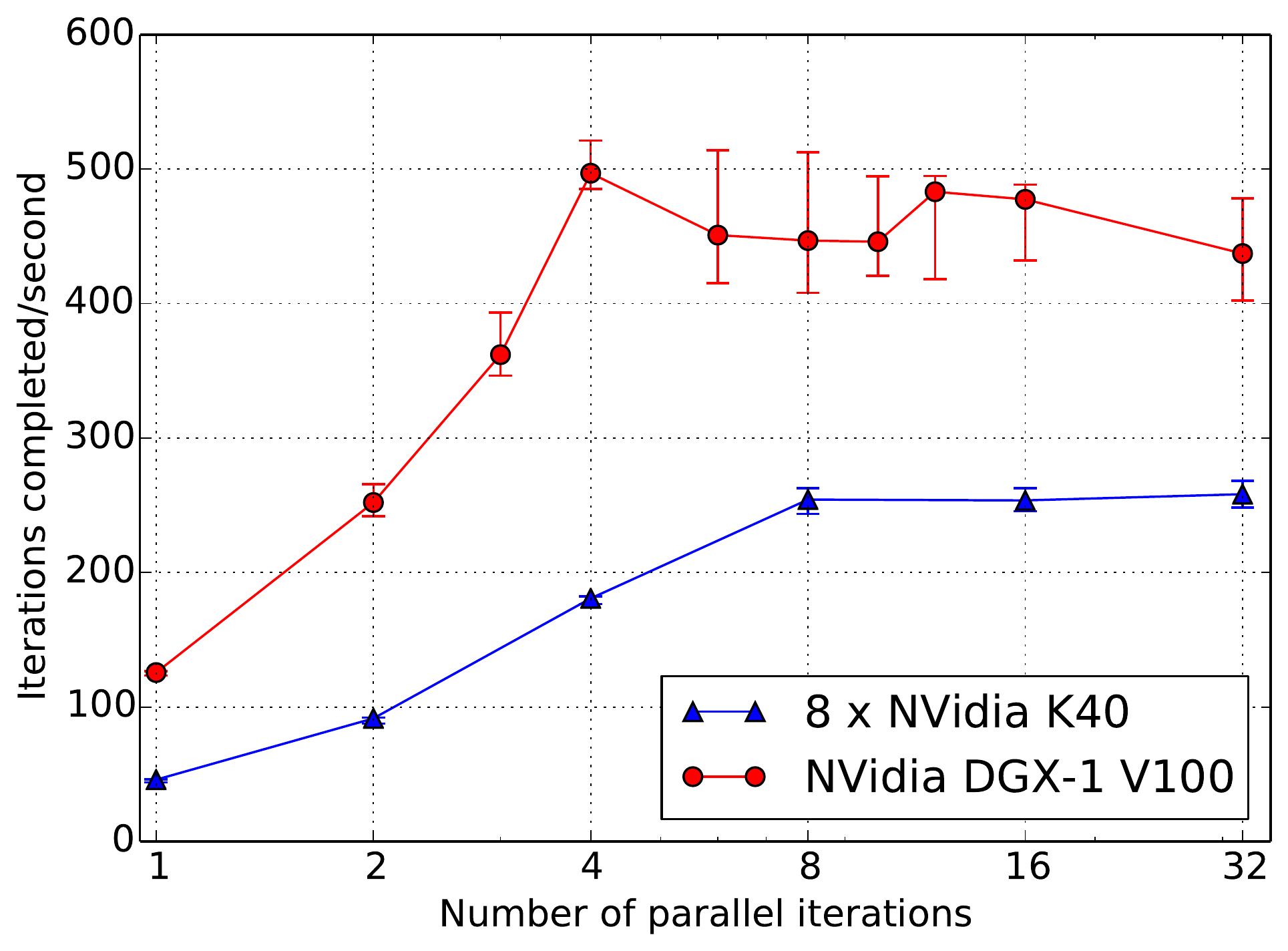}
  \end{center}
  \caption{Effect of changing the number of iterations that are allowed to run concurrently.}\label{fig:eval:parallel_iterations}
  \vspace{-3mm}
\end{figure}

Figure~\ref{fig:eval:parallel_iterations} demonstrates that running iterations
in parallel is crucial for achieving the inherent parallelism from the $8$ GPUs;
computation is automatically pipelined so as to mask any data dependencies.  On
the K40 machine, we reach the peak performance when the setting for parallel
iterations is above~$8$.  On the faster V100 server we achieve highest
performance with $4$ parallel iterations, but additional parallelism introduces
scheduling noise. This experiment also gives us a comparison with out-of-graph
loop execution. When the parallel iteration count is set to $1$, loop
iterations are executed sequentially, similarly to straightforward out-of-graph
execution driven by a single thread. As
Figure~\ref{fig:eval:parallel_iterations} indicates, in-graph control flow makes
it easy to exploit parallelism, here giving $5$ times more iterations per second
than the unparallelized approach.

\subsection{Memory Management}\label{ss:eval:memory}

In this experiment we evaluate the effectiveness of swapping memory
between CPU and GPU. {\tf}'s RNN implementation (known as \texttt{dynamic\_rnn})
is based on our work, and has been used in a wide variety of machine learning
applications, including Neural Machine Translation
(see Section~\ref{sec:prog}). We use \texttt{dynamic\_rnn} in our experiments.
The model
architecture in this experiment is a single-layer LSTM~\cite{hochreiter1997long}
with 512 units. The RNN implementation is available in the current distribution of {\tf}~\cite{tfv16}.



\begin{table}[b]
  \centering
      {
        \small
        \begin{tabular}{c|rrrrrrr}
          & \multicolumn{7}{c}{Training time per loop iteration (ms), by sequence length}\\
          \cline{2-8}
          Swap & 100 & 200 & 500 & 600 & 700 & 900 & 1000\\
          \hline
          Disabled & 5.81 & 5.78 & 5.75 & OOM & OOM & OOM & OOM\\
          Enabled & 5.76 & 5.76 & 5.73 & 5.72 & 5.77 & 5.74 & 5.74\\
        \end{tabular}
      }
      \caption{Training time per loop iteration for an LSTM model with increasing
        sequence lengths.}\label{tbl:eval:swap}
\vspace{-1.5em}
\end{table}

\begin{figure*}
\begin{center}
  \includegraphics[width=0.8\linewidth]{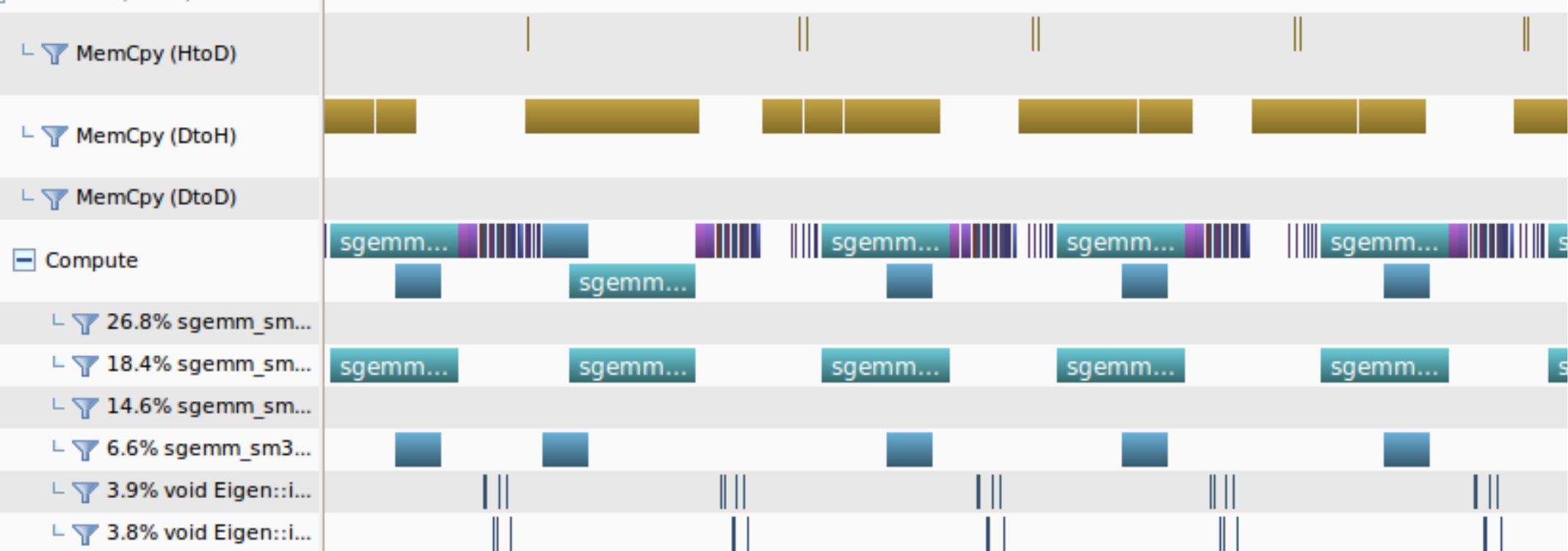}
\end{center}
\caption{Timelines for the GPU kernels with memory swapping enabled.}\label{fig:eval:timeline}
\end{figure*}

One key measure for RNN performance is the ability to train on long sequences
with large batch sizes. Long sequences are important in many applications;
examples include patient history in health care, view history in web sites that
recommend content, and signal sequences in speech or music. The use of
sufficiently large batch sizes is crucial for efficiency on devices such as
GPUs. Table~\ref{tbl:eval:swap} shows the performance of training as we increase
the sequence length from $100$ to $1000$. All results are for a single GPU with
batch size $512$. When memory swapping is disabled, we run out of memory (OOM)
at sequences of length a little over $500$. When memory swapping is enabled, we
can train on sequences of length $1000$ with no overhead. This increase in
sequence length allows users to train substantially larger and deeper networks
(e.g., with multiple LSTM layers) at no additional cost.
For those models, the maximum length of a sequence whose state fits in GPU
memory would decrease even further (e.g., to $500 / 8$ for an 8-layer LSTM model)
in the absence of optimizations such as swapping or model parallelism (which we
discuss in Section~\ref{ss:eval:model}).

We attribute this scalability to the ability to overlap
compute operations with the I/O operations for memory swapping.
This ability arises from the combination of
parallel iterations, multi-stream asynchronous GPU kernel execution,
and asynchronous state saving in gradient computations (\S\ref{sec:diff:memory}).
Figure~\ref{fig:eval:timeline} shows the timelines for the kernels executed
on both the compute and I/O streams of the GPU. The kernels on the
compute stream (labeled as Compute) are the compute operations of the LSTMs; the
kernels on the two I/O MemCpy streams (labeled as DtoH and HtoD) are copy kernels that transfer tensors
between CPU and GPU. The figure shows a time window from the
forward computation, so the GPU-to-CPU stream (DtoH) is active and the
CPU-to-GPU stream (HtoD) is mostly idle.
As the figure indicates, the execution of the compute kernels and the I/O
kernels proceed in parallel, so the total elapsed time
with memory swapping is almost identical to the elapsed time without it.

\subsection{Dynamic Control Flow vs.~Static Unrolling}\label{ss:eval:static}

\begin{figure}
  \begin{center}
    \includegraphics[width=\linewidth]{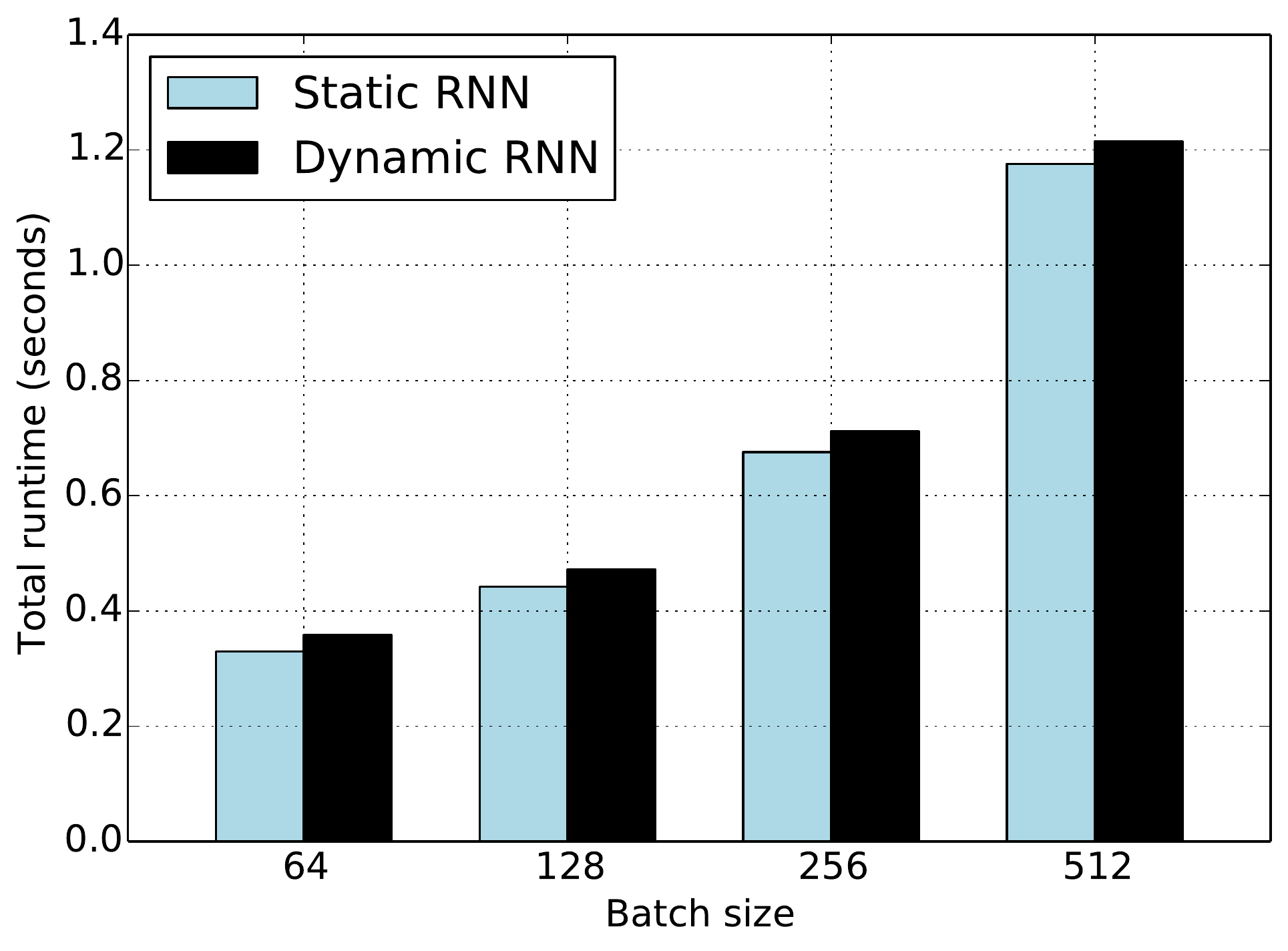}
  \end{center}
  \caption{Performance comparison of dynamic control flow against
    static unrolling.}\label{fig:eval:static}
\end{figure}


An alternative to using \texttt{dynamic\_rnn} is to rely on static
loop unrolling.  Static unrolling eliminates dynamic control flow and
therefore gives a good baseline in understanding the performance of
dynamic control flow.

Figure~\ref{fig:eval:static} shows the total elapsed times of running
one training step with various batch sizes. All results are for a
single-layer LSTM running on one GPU with sequence length $200$. We
see a small slowdown of between $3\%$ and $8\%$, and the slowdown decreases
as we increase the batch size (and hence the computation). The slowdown
is largely due to the overhead of dynamic control flow.

We also consider the memory consumption of \texttt{dynamic\_rnn}
against static unrolling. In some configurations,
\texttt{dynamic\_rnn} can handle substantially longer sequences than
static unrolling. For example, for a single layer LSTM model of $2048$
units and batch size $256$, \texttt{dynamic\_rnn} can handle sequences
of length $256$ while static unrolling runs out of memory at $128$.
Static unrolling exposes the entire unrolled dataflow graph (and hence
all the potential parallelism) to the runtime. The abundant
parallelism poses a challenge to the runtime as the
order of the operations can dramatically impact memory usage. With dynamic
control flow, the additional structure
enables the runtime to choose an execution order that is both
time- and memory-efficient.


%
%

\subsection{Model Parallelism}\label{ss:eval:model}

\begin{figure}
  \begin{center}
    \includegraphics[width=\linewidth]{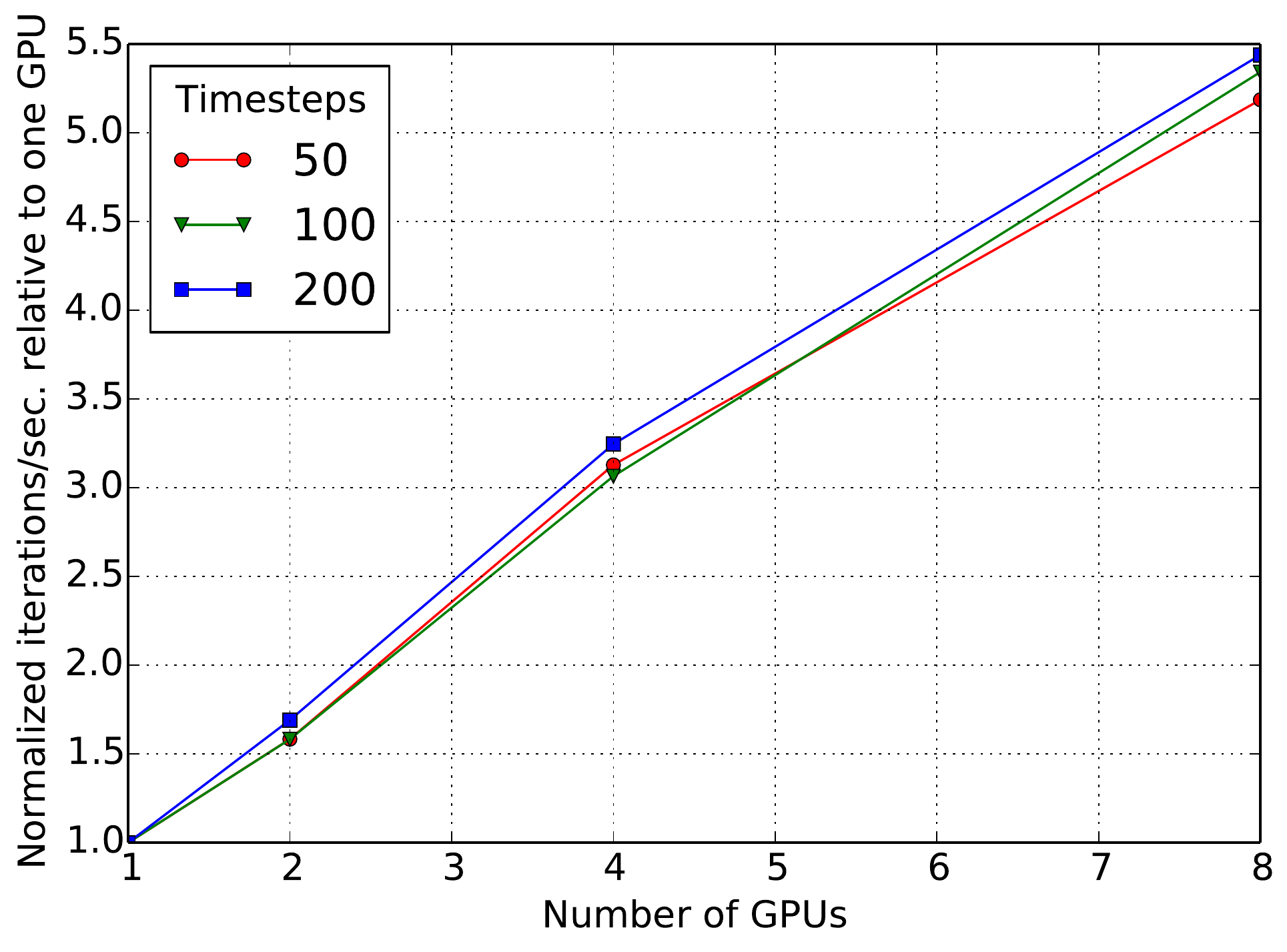}
  \end{center}
  \caption{Parallel speedup for an $8$-layer LSTM as we vary the
    number of GPUs from 1 to 8.}\label{fig:eval:model_parallel}
\end{figure}

For several reasons such as better memory utilization, users often want to
train models in parallel across multiple devices. For example, a multi-layer
RNN can be parallelized across multiple GPUs by assigning each layer to a different
GPU. In our approach, this strategy is expressed as a single loop that is
partitioned across GPUs.
Recall that in Figure~\ref{fig:eval:parallel_iterations}, we show how a
microbenchmark performs in such a setting. We now evaluate the performance
of an end-to-end training step of a realistic $8$-layer LSTM model on 8 GPUs.
Figure~\ref{fig:eval:model_parallel} shows the total elapsed time as we vary
the number of GPUs. We observe a parallel speedup from $1$ to $8$ GPUs,
and a speedup of $5.5\times$ at $8$ GPUs. As expected, the speedup is sub-linear,
due to the additional DMA overhead when using multiple GPUs, but this is mitigated
by the ability to overlap computation in multiple iterations.
The running time includes
the gradient computation, so this experiment additionally illustrates the performance of the
parallel and distributed execution of gradient computations.

\subsection{An Application: Reinforcement Learning}\label{ss:eval:rl}

\begin{figure}
\begin{center}
  \includegraphics[width=.9\linewidth]{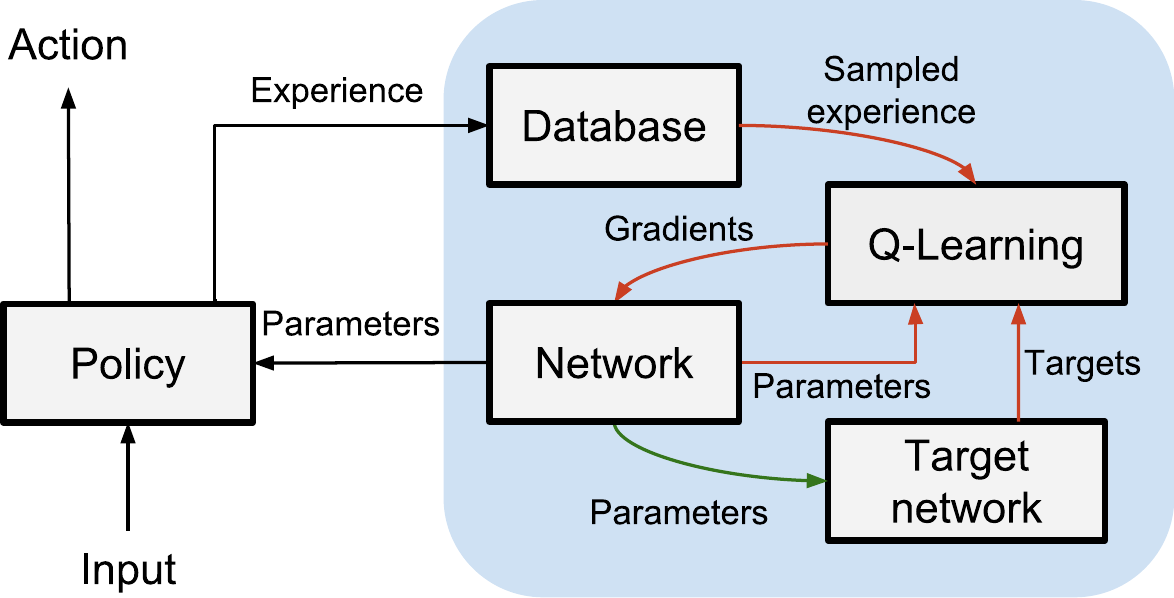}
\end{center}
\caption{Dynamic control flow in Deep Q-Networks.}\label{fig:diff:dqn}
\end{figure}

Finally, we consider the benefits of dynamic control flow in an
archetypical application: we describe how Deep Q-Networks
(DQN) \cite{mnih2015human}, a benchmark reinforcement learning algorithm,
can be implemented with dynamic control flow. While DQN has already been
superseded by newer methods, this example is representative of uses of dynamic
control flow in reinforcement learning, including more recent algorithms.

Figure~\ref{fig:diff:dqn} shows a diagram of the DQN algorithm.
DQN augments a neural network with a database in which it stores its incoming
experiences, periodically sampling past experiences to be employed for
Q-Learning, a form of reinforcement learning. Q-Learning uses these experiences
along with a second neural network, called the target network, to train the
main network. The target network is periodically updated to reflect a snapshot
of the main network.

The DQN algorithm includes many conditionals: different experiences cause
different types of writes to the database, and sampling of experiences,
Q-Learning, and updates to the target network are each performed conditionally
based on the total amount of experience received.


A baseline implementation of DQN without dynamic control flow requires
conditional execution to be driven sequentially from the client program. The
in-graph approach fuses all steps of the DQN algorithm into a single dataflow
graph with dynamic control flow, which is invoked once per interaction with the
reinforcement learning environment. Thus, this approach allows the entire
computation to stay inside the system runtime, and enables parallel execution,
including the overlapping of I/O with other work on a GPU.  It yields a speedup
of $21\%$ over the baseline.  Qualitatively, users report that the in-graph
approach yields a more self-contained and deployable DQN implementation; the
algorithm is encapsulated in the dataflow graph, rather than split between the
dataflow graph and code in the host language.

%
%

%
%
%


%% file: rw.tex
\section{Related Work}\label{sec:rw}

Our approach to control flow draws on a long line of research on
dynamic dataflow architectures, going back to the work of Arvind et
al.~\cite{Arvind:1986,Arvind:1990}. The
timely dataflow model~\cite{murray2016timely}, implemented in the
Naiad system~\cite{murray2013naiad}, can be seen as a recent
embodiment of those architectures.  It supports distributed execution,
with protocols for tracking the progress of computations. The
control-loop state machines we describe in Section~\ref{sec:impl:dist}
constitute a specialized approach for this tracking that is more lightweight and
efficient for the task;
this approach, although suitable for our purposes,
would be difficult to extend to incremental computations
of the kind that Naiad enables. Naiad does
not support heterogeneous systems, so does not address some of
the problems that we studied, in particular memory management across
heterogenous devices. Nor does it
address automatic differentiation, which is crucial for machine
learning applications.


Some systems for machine learning, such as
Theano~\cite{bergstra2010theano,DBLP:journals/corr/abs-1211-5590}
and CNTK~\cite{conf/kdd/SeideA16}, allow the programmer
to create a computation graph, for example with a Python front-end, and
then to launch the execution of this graph, following the in-graph approach described in the introduction.
Theano allows this graph to contain
control-flow constructs, but Theano's support for control flow is relatively limited.
In particular, Theano allows neither
nested loops, nor the parallel or distributed execution of control-flow constructs;
its automatic differentiation often requires more computation and is therefore less
efficient than the approach in this paper. In a discussion of
limitations and challenges related to control flow, the developers of
Theano have written that they find {\tf}'s approach
appealing~\cite{alrfou2016theano}.


Other systems for machine learning, such as
Torch~\cite{collobert2002torch}, Chainer \cite{chainerorg}, and
PyTorch~\cite{pytorch}, blur this phase distinction: a graph appears
to be executed as it is defined, either on one input example at a time or on
manually specified batches.  Because the graph is not given ahead of
time, optimizations are more difficult. These systems are typically
based on Python, and expose Python's control-flow constructs, which do
not come with support for distributed execution and memory management
across heterogeneous devices.  The two approaches are reconciled in
systems such as MXNet~\cite{chen2015mxnet,mxnet-github}, which supports
both (but without control flow in graphs), and with {\tf}'s
``imperative mode'' and ``eager mode''
extensions~\cite{manjunath,eager-blogpost}.

While we favor embedding control flow in a static graph, others have
proposed more dynamic distributed execution engines that support
similar control flow. For example, CIEL represents a program as an
unstructured ``dynamic task graph'' in which tasks can
tail-recursively spawn other tasks, and imperative control-flow
constructs are transformed into continuation-passing
style~\cite{murray2011ciel}. Nishihara et al.~recently
described a system for ``real-time machine learning'' that builds on
these ideas, and adds a decentralized and hierarchical scheduler to
improve the latency of task
dispatch~\cite{nishihara2017ray}. Programming models based on dynamic
task graphs are a direct fit for algorithms that make recursive
traversals over dynamic data structures, such as parse trees~\cite{goller1996bpts}.
By contrast, in our approach this
recursion must be transformed into iteration, for example using the
transformation that Looks et al.~describe~\cite{DBLP:journals/corr/LooksHHN17}.
The drawback of an
unstructured dynamic task graph is that the individual tasks
are black boxes, and more challenging to optimize holistically.

The wish to save memory is fairly pervasive across machine learning
systems, in part because of the important role of GPUs and other
devices with memory systems of moderate size. Some techniques based on
recomputation specifically target the memory requirements of
backpropagation~\cite{DBLP:journals/corr/ChenXZG16,DBLP:journals/corr/GruslysMDLG16}. These
techniques concern feed-forward graphs, LSTMs, and RNNs, rather than
arbitrary graphs with control-flow constructs; for particular classes
of graphs, clever algorithms yield efficient recomputation policies,
sometimes optimal ones.  Our work on swapping belongs in this line of
research.  So far, we have emphasized the development of mechanisms,
and relatively simple but effective policies for their use. In future
work we may explore additional algorithms and also the application of
reinforcement learning to swapping and recomputation decisions.

Finally, our research is related to a substantial body of work on
automatic differentiation
(e.g.,~\cite{TapenadeRef13,DBLP:journals/toplas/PearlmutterS08,autograd}).
That work includes systems for automatic
differentiation of ``ordinary'' programming languages (e.g., Fortran,
C, or Python) with control-flow constructs.  
It has generally not been concerned
with parallel and distributed implementations---perhaps because working
efficiently on large datasets, as in deep learning, has not been a common goal~\cite{DBLP:journals/corr/BaydinPS16a}.

%% file: conc.tex
\section{Conclusions}\label{sec:conc}

This paper presents a programming model for machine learning that
includes dynamic control flow, and an implementation of that model.
This design enables parallel and distributed execution in
heterogeneous environments. Automatic differentiation contributes to
the usability of the control-flow constructs.  The implementation
allows us to leverage GPUs and custom ASICs with limited memory, and
supports applications that make frequent control-flow decisions across
many devices. Our code is part of {\tf}; it has already
been used widely, and has contributed to new advances in machine learning.

Dynamic control flow relates to active areas of machine learning,
which may suggest opportunities for further work. In particular,
conditional computation, where parts of a neural network are active on
a per-example basis, has been proposed as a way to increase model
capacity without a proportional increase in computation; recent
research has demonstrated architectures with over 100 billion
parameters~\cite{DBLP:journals/corr/ShazeerMMDLHD17}.  Also,
continuous training and inference may rely on streaming systems, perhaps using
the dynamic dataflow architectures that we adapt, as
suggested by work on timely dataflow and differential
dataflow~\cite{murray2016timely}.  Further research on 
abstractions and implementation techniques for conditional and
streaming computation seems worthwhile.

Dynamic control flow is an important part of bigger trends that we have
begun to see in machine learning systems. Control-flow constructs
contribute to the programmability of these systems, and enlarge the set of models that
are practical to train using distributed resources.  Going further, we
envision that additional programming-language facilities will be
beneficial. For instance, these may include abstraction mechanisms and support
for user-defined data structures. The resulting design and
implementation challenges are starting to become clear.
New compilers and run-time systems, such as XLA~\cite{xla}, will undoubtedly play a role.




